\documentclass[prd,notitlepage,nofootinbib,superscriptaddress]{revtex4-1}

\usepackage{comment}
\usepackage{mathrsfs}
\usepackage{yfonts}
\usepackage{tipa}
\usepackage{bm}
\usepackage{bbm}
\usepackage{amsmath}
\usepackage{amssymb}
\usepackage{amsthm}
\usepackage{amsfonts}
\usepackage{mathtools}
\usepackage{accents}
\usepackage{color}
\usepackage{latexsym}
\usepackage{relsize}
\usepackage[greek,english]{babel}
\usepackage{booktabs}
\usepackage[iso-8859-7]{inputenc}
\usepackage[colorlinks=true]{hyperref}
\usepackage[vcentermath]{youngtab}
\usepackage{ytableau}
\usepackage{tikz}
\usetikzlibrary{shapes}
\usetikzlibrary{positioning}
\usepackage{soul}
\usepackage{array}
\newcolumntype{?}{!{\vrule width 2pt}}

\newcommand{\lp}{\left(}
\newcommand{\rp}{\right)}
\newcommand{\lb}{\left[}
\newcommand{\rb}{\right]}

\newcommand{\bX}{\boldsymbol{\mathrm{X}}}
\newcommand{\bH}{\boldsymbol{\mathrm{H}}}

\newcommand{\lsim}   {\mathrel{\mathop{\kern 0pt \rlap
  {\raise.2ex\hbox{$<$}}}
  \lower.9ex\hbox{\kern-.190em $\sim$}}}
\newcommand{\gsim}   {\mathrel{\mathop{\kern 0pt \rlap
  {\raise.2ex\hbox{$>$}}}
  \lower.9ex\hbox{\kern-.190em $\sim$}}}
\newcommand{\bw}{\begin{widetext}\begin{equation}}
\newcommand{\ew}{\end{equation}\end{widetext}}
\newcommand{\be}{\begin{equation}}
\newcommand{\ee}{\end{equation}}
\newcommand{\ba}{\begin{eqnarray}}
\newcommand{\ea}{\end{eqnarray}}

\newcommand{\diff}{{{\rm d}}}

\newcommand{\Diff}{\mathrm{D}}

\newcommand{\bt}{\mathbf{t}}

\newcommand{\bxi}{\boldsymbol{{\xi}}}

\newcommand{\e}{\mathrm{e}}
\newcommand{\ie}{\text{\textschwa}}

\newcommand{\bL}{\boldsymbol{\mathrm{L}}}
\newcommand{\bJ}{\boldsymbol{\mathrm{J}}}

\newcommand{\bT}{\mathbf{T}}

\newcommand{\bP}{\boldsymbol{\mathrm{P}}}
\newcommand{\bA}{\mathbf{A}}
\newcommand{\bF}{\mathbf{F}}

\newcommand{\bLambda}{\mathbf{\Lambda}}

\newcommand{\bQ}{\mathbf{Q}}

\newcommand{\nn}{\nonumber}

\begin{document}

\title{An axiomatic purification of gravity}

\author{Tomi Koivisto}
\email{t.s.koivisto@astro.uio.no}
\affiliation{Laboratory of Theoretical Physics, Institute of Physics, University of Tartu, W. Ostwaldi 1, 50411 Tartu, Estonia}
\affiliation{National Institute of Chemical Physics and Biophysics, R\"avala pst. 10, 10143 Tallinn, Estonia}
\author{Manuel Hohmann}
\email{manuel.hohmann@ut.ee}
\affiliation{Laboratory of Theoretical Physics, Institute of Physics, University of Tartu, W. Ostwaldi 1, 50411 Tartu, Estonia}
\author{Luca Marzola}
\email{luca.marzola@cern.ch}
\affiliation{National Institute of Chemical Physics and Biophysics, R\"avala pst. 10, 10143 Tallinn, Estonia}

\date{\today}

\begin{abstract}
A theory of gravity is deduced from the axioms of the premetric program. The starting point is the conservation of energy and momenta,
and the equivalence of gravitation and inertia.  The latter is what leads to the framework of the so called purified gravity. The local and linear
constitutive relation has 14 components when it is assumed to be metrical, but the compatibility of the constitutive relation with an action principle
fixes uniquely the theory of Coincident General Relativity. The premetric formalism of purified gravity has a direct analogy with
massive electromagnetism, the Planck mass corresponding to the Proca mass of the gauge boson. The metric emerges as a Stueckelberg field, and the graviton as a
Goldstone boson of the broken symmetry.
\end{abstract}


\maketitle


\tableofcontents


\section{Introduction}

Electrodynamics, in Maxwell's well-known form and its many possible generalisations, can be understood to a surprising extent without referring to a metric. The starting point is then to consider quantities that can be counted, without requiring the measurements
of areas, volumes or durations, for which one typically would need to resort to less elementary objects such as measuring sticks and clocks. The invariance of the countable elementary quantities, in particular electric charges, gives rise to a conserved current (3-form), and this in turn gives rise to an excitation (2-form), whereas the conservation of the magnetic flux gives rise to a field strength (2-form). The workings of theory are then specified by the relation between the field strength and the excitation, called the constitutive relation. In Maxwell's vacuum electrodynamics the constitutive relation (Hodge dual) requires a metric, but more general possibilities can be considered, with or without invoking a metric, and this allows the unified
description of the vast variety of physical phenomenology of the electromagnetic interaction from linear and non-linear effects in media to axionic and other extensions of the Maxwell electrodynamics.

Such a {\it premetric} construction of the classical electromagnetic theory is exposed in great detail and clarity in the pedagogical textbook of Hehl and  Obukhov \cite{hehl2003foundations}. The premetric program was originally put forward in 1922 by Kottler, who applied it both to
electromagnetism and to Newtonian gravity \cite{Hehl:2016glb}. More recently, relativistic theories of gravity have been considered in the context of the premetric program \cite{Itin:2016nxk,Hohmann:2017duq,Itin:2018lcb,Itin:2018dru,Puetzfeld:2019wwo,Obukhov:2019oar}, and this has quite naturally lead to the metric teleparallel\footnote{A flat affine connection is called  ``teleparallel''. We refer to a metric-compatible flat connection as  ``metric teleparallel'',
and to symmetric flat connection as  ``symmetric teleparallel''. (In the literature the term teleparallel usually means the former special case, and it is a common statement that whereas in General Relativity the fundamental variable is the metric, in the teleparallel version of the theory the fundamental variable is the tetrad. However, such a statement is empty, if not misleading, since the Einstein-Hilbert action can just as well be rewritten in solely terms of the tetrad. A unifying framework for all the formulations is metric-affine theory, wherein the fundamental variables are the metric and the affine connection \cite{BeltranJimenez:2019tjy,Jimenez:2019ghw}. In General Relativity the appropriate connection of course is not teleparallel, but it is both metric and symmetric.) } reformulation of Einstein's theory \cite{Aldrovandi:2013wha,Maluf:2013gaa}. In the premetric approach to the theory of gravity, one begins with the conservation of energy and momenta since these are the sources of the gravitational field \cite{Hehl:2019csx}. Formally, the construction proceeds in a rather direct analogy to
the case of electromagnetism. The conservation of energy and momenta gives rise to currents and further, excitations. Corresponding forces are introduced, and now their constitutive relation to the excitations, even in the local and linear case, contains many more possibilities than
in the case of electromagnetism \cite{Itin:2018dru}, due to there being four conserved charges instead of one.

In the standard textbook descriptions of General Relativity, gravitation is often interpreted as geometry \cite{Misner:1974qy}, whereas
in the metric teleparallel formulation gravity is rather understood as a force \cite{Aldrovandi:2013wha}. These alternative formulations and interpretations provide
interesting insights into both Newton's and Einstein's theories \cite{KNOX2011264}, but yet, it may be also useful to recall that to the latter, the quintessence of gravitation was neither
force nor geometry, but inertia \cite{pittphilsci9825}. To express this idea precisely, in the modern terms of a gauge theory, one may contemplate the basic fact that a purely inertial interaction
 is characterised by a vanishing gauge field strength. It is the gauge field strength that is both the gauge invariant measure of force, from the mathematical perspective of field theory, and the gauge invariant measure of geometry, from the perspective of principal bundles. However, the description has to be a bit more subtle, since the gravitational force (or, equivalently, geometry) can be eliminated only locally.

A resolution for the dilemma has been recently sought in the context of the so called {\it purified gravity} \cite{BeltranJimenez:2017tkd}. From the viewpoint of geometrical foundations, it was proposed that the fictitious forces could be described by a purely integrable spacetime affine geometry \cite{Koivisto:2018aip}. Such a geometry, which is devoid of both torsion and curvature, was recognised as that of the so called symmetric teleparallelism \cite{Nester:1998mp,Adak:2008gd}  and it was discovered that the affine connection of the geometry is generated by a pure coordinate transformation, i.e. a (passive) translation. This was the starting point for a reformulation of Einstein's theory, called the Coincident General Relativity (CGR) \cite{BeltranJimenez:2017tkd}. This theory, which is determined uniquely by the integrability postulate and a symmetry principle, can indeed be understood as a canonical gauge theory of translations, and it can be simply described as the minimal covariantisation of the Einstein's action \cite{Jimenez:2019yyx}. For a review of the three possible formulations of General Relativity, in terms of curvature, torsion and non-metricity, respectively, see Ref. \cite{BeltranJimenez:2019tjy}, for a unification of teleparallel geometries see the recent Ref. \cite{Jimenez:2019ghw}, and about more general modifications of General Relativity, see e.g. Ref. \cite{Heisenberg:2018vsk}.

In this paper, we attempt to understand purified gravity from the perspective of the premetric program. It will turn out that the electromagnetic analogy to purified gravity is rather Proca's massive
\cite{Ruegg:2003ps,Tu:2005ge,Goldhaber:2008xy} than Maxwell's massless electromagnetism. This clarifies why also the
gravitational field itself contributes to the conserved currents of energy and momenta (which is more difficult to explain in the metric teleparallel construction), and the dimension of the gravitational action is no longer anomalous. A main conclusion will be that in the consistent
formulation of the theory, the field strength is vanishing but there exists an excitation: this reflects the geometrical set-up wherein the affine spacetime connection is trivialisable, but the connection to which matter turns out to couple,
is the curved metric-compatible connection \cite{Koivisto:2018aip}. Actually, the more suitable analogy would be Stueckelberg's than Proca's massive electromagnetism. The latter's formulation is physically equivalent but respects the original symmetry of the Maxwell theory
(which, though coming at the price of introducing an extra scalar, can be indispensable for e.g. renormalisation \cite{Ruegg:2003ps}). Indeed, the constitutive relation of CGR is found to be uniquely specified as the one that
restores the translational symmetry of the theory. Thus, the graviton can be interpreted as the massless Goldstone boson of a spontaneous symmetry breaking, an idea which goes back to at least to Isham, Salam and Strathdee \cite{Isham:1971dv}.

The structure of this article is seen from the Table of contents above. First we shall go through the premetric deduction of gravity theory using the standard language of tensors in Section \ref{tensors}. It can be helpful, in the spirit of \cite{BeltranJimenez:2018vdo}, to expose the basic foundation of the construction without obscuring its simplicity by excessive mathematical formalism. On the other hand, exterior algebra provides the natural expressions for conservation laws, and elegantly highlights the paramount role of the Poincar\'e's lemma in the premetric reasoning. Thus in Section \ref{forms} we also write down the premetric construction in the language
of differential forms. The reader familiar with this language might prefer to start from Section \ref{forms}, Section \ref{tensors} being redundant with it. In both discussions, we emphasise the analogy between gravity and electromagnetism by first reviewing the premetric perspective in the latter, slightly simpler, case (a comparison of these two cases and a dictionary between the two languages will then be given in Table \ref{table1}).
The local and linear constitutive relations are analysed in Section \ref{constitutions}, by first focusing on metrical relations, taking into account parity-violating ones, and then generalising to fully arbitrary constitutive relations which we decompose into their irreducible components. The interpretation of the metric as a Stueckelberg field is elaborated in Section \ref{symmetries}, filling in some technical details and briefly speculating on the ultraviolet limit of purified gravity. The properties of the theories with more general constitutive relations, that do not share the unique property of the CGR relation, are explored in Section \ref{properties}, with attention on the degrees of freedom, propagation of waves and the conservation laws. We conclude in Section \ref{conclusions} with a summary and discussions.

\section{Premetric construction in the tensor language}
\label{tensors}

We shall refer to the covariant derivative that satisfies $[\nabla_\mu,\nabla_\nu]=0$. We ask the reader who is uncomfortable with this to kindly just consider $\nabla_\mu$ as a notation for $\partial_\mu$
until the Section \ref{ptos}, wherein we shall justify the use of the covariant form of the equations in this Section.

\subsection{Excitation}

\subsubsection{Electromagnetism}

The conservation of electric charge entails the existence of an electric current, described by the vector density $J^\mu$.  The charge conservation, in integral and in differential forms is
\be
\int_{\partial \Omega_4}\diff^3 x J^\mu n_\mu =0\,, \quad \text{and} \quad \nabla_\mu J^\mu = 0\,, \label{chargecons}
\ee
respectively, $n_\mu$ being a unit normal to the 3-surface $\partial\Omega_4$. Locally, this is equivalent to the equation (that is a generalised version of the inhomogeneous Maxwell equation)
\be \label{inhomomax}
\nabla_\mu H^{\mu\nu} = J^\nu\,,
\ee
where the electromagnetic excitation $H^{\mu\nu}=H^{[\mu\nu]}$ is an antisymmetric tensor density. In case of a theory with self-interactions, such as in Proca's massive electromagnetic theory
or a non-Abelian gauge theory,  we may write $J^\mu=T^\mu + t^\mu$, where $T^\mu$ are the external sources and $t^\mu$ are due to the electromagnetic field
itself\footnote{To give a concrete example, in the Proca theory we would have $t^\mu = m^2\sqrt{-g} g^{\mu\nu} A_\nu$, where $m$ is the mass of the electromagnetic field, $g^{\mu\nu}$ is the metric and  $A_\nu$ is the electromagnetic potential we shall introduce in a moment. At this point of course we do not have a metric at hand.}. We have assumed an additive decomposition of the
sources. There is redundancy in the excitation tensor density, in the sense that any  $H^{\mu\nu} \rightarrow H^{\mu\nu} + \nabla_\lambda \varphi^{\lambda\mu\nu}$, where $\varphi^{\lambda\mu\nu}{}= \varphi^{[\lambda\mu\nu]}$ also satisfies the above equation (\ref{inhomomax}) with the antisymmetric property of as $H^{\mu\nu}$. Without requiring the same property, the 4-component redundancy is increased to 24 components. This ambiguity of the excitation tensor is not usually taken into account in the premetric construction of electromagnetic theory, since it has no relevance to the dynamics.

\subsubsection{Gravity}

In gravity, we begin with the conservation of energy-momentum and denote the corresponding current as $J^{\mu}{}_{\nu}$.
The four conservation laws are, again in integral and in differential forms, expressed as
\be
\int_{\partial \Omega_4}\diff^3 x J^\mu{}_\nu n_\mu =0\,, \quad \text{and} \quad \nabla_\mu J^\mu{}_\nu = 0\,.
\ee
The latter implies again the existence of an antisymmetric excitation tensor density $H^{\mu\nu}{}_\alpha = H^{[\mu\nu]}{}_\alpha$.
The redundancy in this tensor density is $H^{\mu\nu}{}_\alpha \rightarrow H^{\mu\nu}{}_\alpha + \nabla_\lambda \varphi^{\lambda\mu\nu}{}_\alpha$ where, if  $\varphi^{\lambda\mu\nu}{}_\alpha = \varphi^{[\lambda\mu\nu]}{}_\alpha$ it has 16 independent components, and if not, 96 independent components.
We now write
\be \label{gravity}
\nabla_\mu H^{\mu\nu}{}_\alpha = J^\nu{}_\alpha = T^\nu{}_\alpha + t^\nu{}_\alpha\,,
\ee
taking into account that in addition to the energy-momentum of matter $T^\mu{}_\nu$, there can also occur inertial energy-momentum $t^\mu{}_\nu$.

\subsection{Field strength}

\subsubsection{Electromagnetism}

The field strength $F_{\mu\nu}$ satisfies the integral and differential conservation equations
\be
\int_{\partial \Omega_4}F_{\mu\nu}n^\nu =0\,, \quad \text{and} \quad \nabla_{[\alpha}F_{\mu\nu]} = 0\,,
\ee
implying the conservation of the magnetic flux and the existence of the electromagnetic potential
$A_\mu$, such that $F_{\mu\nu} = 2\nabla_{[\mu}A_{\nu]}$. It is defined up to the total derivative $A_\mu \rightarrow A_\mu  + \nabla_{\mu}\varphi$. For a detailed premetric investigation of the potential $A_\mu$, see \cite{Pfeifer:2016har}.

\subsubsection{Gravity}

In analogy to electromagnetism, we introduce the gravitational field strength $F^{\alpha\beta}{}_{\mu\nu}$ which ought to be conserved,
\be
\int_{\partial \Omega_4}F^{\alpha\beta}{}_{\mu\nu}n^\nu =0\,, \quad \text{and} \quad \nabla_{[\rho}F^{\alpha\beta}{}_{\mu\nu]} = 0\,.
\ee
This implies the existence of a gravitational potential $A^{\alpha\beta}{}_{\mu}$, such that $F^{\alpha\beta}{}_{\mu\nu} = 2\nabla_{[\mu}A^{\alpha\beta}{}_{\nu]}$ and
defined up to the derivative $A^{\alpha\beta}{}_\mu \rightarrow A^{\alpha\beta}{}_\mu + \nabla_{\mu}\varphi^{\alpha\beta}$. The defining peculiarity of gravitation is that it is can be
always be locally eliminated. In other words, its field strength should vanish, $F^{\alpha\beta}{}_{\mu\nu} = 0$. Therefore, we can always assume that $A^{\alpha\beta}{}_\mu = \nabla_{\mu}\varphi^{\alpha\beta}$.

One notes that we stipulated that the gravitational field strength comes with two indices, thus the transformation of $\varphi^{\mu\nu}$ has a priori 16 independent components.
Later, when imposing the constitutive relations, it will be evident that the theory actually involves only the 10 components that are symmetric with respect to the exchange of the two indices. A possible interpretation
is that these correspond to the 4+6 conserved quantities, the four-momentum and the angular momenta. One may also consider that the underlying symmetry is just the symmetry of the frame, GL(4), and by requiring Lorentz invariance (through the imposement of the symmetrised constitutive relations) we can then eliminate the 6 antisymmetric components of $\varphi^{\alpha\beta}$, leaving us with the 10 nonzero $\varphi^{\alpha\beta}=\varphi^{(\alpha\beta)}$.

We can already anticipate that is possible to identify the gauge potential and the pure gauge transformation as $A^{\alpha\beta}{}_\mu = -Q_\mu{}^{\alpha\beta}$ and $\varphi^{\alpha\beta}=g^{\alpha\beta}$, respectively. The vanishing of the
field strength means teleparallelism, $\nabla_{[\mu}Q_{\nu]}{}^{\alpha\beta} = 0$. It is a geometric identity that
$\nabla_{[\mu}Q_{\nu]}{}^{\alpha\beta} = R^{(\alpha\beta)}{}_{\nu\mu} - \frac{1}{2}T^\lambda{}_{\mu\nu} Q_{\lambda}{}^{\alpha\beta}$, see \cite{BeltranJimenez:2018vdo} for notation and details. However, we should make clear that at this point we do not have a metric at hand. Also, we are not considering a GL(4) gauge theory, where
$R^{(\alpha\beta)}{}_{\nu\mu} \neq F^{\alpha\beta}{}_{\mu\nu}$ would have a different form comprising terms that are quadratic in $A^{\alpha\beta}{}_\mu$ and the gauge transformation (at a nonlinear order) would not be simply the shift by a derivative of $\varphi^{\alpha\beta}$. After arriving at the final form of the premetric theory, we will better clarify its relationship to the theory derived from the conventional gauge approach based on the GL(4) group. Namely, in Section \ref{clarity} we will assume the gauge connection
to be given GL(4) form, denoted by $\Gamma^\alpha{}_{\mu\nu}$ and consisting of Levi-Civita, torsion and non-metricity when such a decomposition is possible, and will then show that in the end the final form of the theory is the same as follows from the present premetric construction.

\subsection{Force}

\subsubsection{Electromagnetism}

The Lorentz force is described by the four-vector density
\be \label{lorentz}
f_\mu = F_{\mu\nu}J^\nu = F_{\mu\nu}\lp T^\mu + t^\mu\rp\,,
\ee
which contributes to the non-conservation of the energy-momentum, as we will learn below.

For the analogous case of gravity, it will be important to realise that in massive electromagnetism there is an element of non-conservation even in the case of vanishing force. Namely, when
we go back to our very starting point (\ref{chargecons}), and recall that in the case of Proca theory in Minkowski space we have $t^\mu=m^2 A^\mu$, the conservation of electric charge current $T^\mu$ is
\be
\nabla_\mu T^\mu = -m^2\nabla_\mu A^\mu\,. \label{electrocons}
\ee
Hence, the electric charge current is conserved only if the Lorentz condition $\nabla_\mu A^\mu =0$ is imposed\footnote{In the case of Proca theory with a curved metric,
the Lorentz condition is generalised to the metric-covariant form and the equation (\ref{electrocons}) becomes the metric-covariant conservation of the electric current tensor (not density).}.

\subsubsection{Gravity}
\label{gforces}

In gravity, the  quantity analogous to (\ref{lorentz}) is a rank (1,1) tensor density
\be
f^{\mu}{}_{\nu} = F^{\mu\beta}{}_{\alpha\nu} \lp T^\alpha{}_\beta + t^\alpha{}_\beta\rp = 0\,, \quad \text{since} \quad F^{\mu\beta}{}_{\alpha\nu} = 0\,.
\ee
The vanishing of the tensor density $f^\mu{}_\nu$ reflects the conservation of the total energy-momentum, to be defined next.

Before that, let us however note that there nevertheless arises an effective force felt by the matter fields. Denoting this effective force by $\mathcal{F}_\nu$, we see that it is given as
\be
\nabla_\mu T^\mu{}_\nu = -\nabla_\mu t^\mu{}_\nu \equiv \mathcal{F}_\nu\,.
\ee
As will be clarified in the following, the presence of this effective force is due to the nonzero mass of the pure gauge connection. That makes it clear that the physical origin of $\mathcal{F}_\nu$
is inertia, though in the end its effects can also be discussed in terms of an effective force and illustrated in terms of geometry. As in the case of electromagnetism, compare (\ref{electrocons}),
the non-conservation is gauge-dependent. The canonical frame of purified gravity \cite{Jimenez:2019yyx} has been proposed to be the analogue of the Lorentz gauge in massive electromagnetism in that, in a well-defined
sense, it establishes the ``physical'' geometry.

\subsection{Energy-momentum current}

\subsubsection{Electromagnetism}

In electromagnetism, the energy-momentum current has the form
\be \label{emt}
\prescript{\text{em}}{}{T}^\mu{}_\nu = H^{\mu\alpha}F_{\nu\alpha} - \frac{1}{4}\delta^\mu_\nu H^{\alpha\beta}F_{\alpha\beta} + P^\mu A_\nu - \frac{1}{2}\delta^\mu_\nu P^\alpha A_\alpha\,,
\ee
where $P^\mu$ is determined by the interaction potential, and its presence generically breaks the U(1) invariance\footnote{An example is the Proca field for which $P^\mu = m^2\sqrt{-g} g^{\mu\nu}A_\nu$.}.
If $P^\mu=0$, the divergence of the above tensor density becomes
\be
\nabla_\mu \prescript{\text{em}}{}{T}^\mu{}_\nu = f_\nu - H^{\alpha\beta}\lp \nabla_\beta F_{\nu\beta} + \frac{1}{4}\nabla_\nu F_{\alpha\beta}\rp - \frac{1}{4}\lp \nabla_\nu H^{\alpha\beta}\rp F_{\alpha\beta}\,.
\ee
It is easy to see that when $H^{\alpha\beta} = F^{\alpha\beta}$, we recover $\nabla_\mu \prescript{\text{em}}{}{T}^\mu{}_\nu = f_\nu$.

\subsubsection{Gravity}

We are considering the purely inertial energy-momentum $t^\mu{}_\nu$ in (\ref{gravity}), since the $T^\mu{}_\nu$ shall be defined by all matter fields.
For example, $\prescript{\text{em}}{}{T}^\mu{}_\nu$ in (\ref{emt}) is a contribution to the $T^\mu{}_\nu$ in the presence of the electromagnetic field, and if there were charged fields contributing to the electromagnetic current
$T^\mu$, they would contribute to the gravitational current $T^\mu{}_\nu$ as well.
Since $F^{\alpha\beta}{}_{\mu\nu}=0$, the kinetic energy-momentum term in direct analogy to (\ref{emt}) is now trivial, but there may be a nontrivial potential energy-momentum.
We denote with $P^\alpha{}_{\mu\nu}$ the conjugate of the potential analogous to $P^\mu$ in electromagnetism. Then, the energy-momentum current analogous to (\ref{emt}) is
\be \label{gemt}
t^\mu{}_\nu = P^\mu{}_{\alpha\beta}Q_{\nu}{}^{\alpha\beta} - \frac{1}{2}\delta^\mu_\nu P^\gamma{}_{\alpha\beta}Q_{\gamma}{}^{\alpha\beta}\,.
\ee
According to Ref. \cite{Jimenez:2019yyx}, this tensor density describes the fictitious energy-momentum in a non-inertial frame, being sourced by a pure-gauge field $A^{\alpha\beta}{}_\mu = \nabla_\mu \varphi^{\alpha\beta} = -Q_\mu{}^{\alpha\beta}$.
The most direct equivalent of this in an electromagnetic theory would the requirement of the vanishing of the current $t^\mu$ in a pure-gauge massive electromagnetism.
It is also trivial to see that the contribution to $\prescript{\text{em}}{}{T}^\mu{}_\nu$ due to a mass of the gauge field $A_\mu$ can be nonzero even in the pure-gauge case $A_\mu=\nabla_\mu\varphi$, but nevertheless it can be always eliminated by choosing the unitary gauge, $\varphi \rightarrow 0$. Whereas formally the equivalent of the Lorentz gauge in a non-trivial Proca theory would be $\nabla_\mu t^\mu{}_\nu = 0$, the definition of the
canonical frame of purified gravity is $t^\mu{}_\nu=0$.

\section{Constitutive relation}
\label{constitutions}

This far we have established the two fundamental equations of purified gravity, expressing the conservation of translational currents and the integrability postulate, respectively.
The quantities appearing in these two equations are related by the linking equations, whose possible forms we study in this Section. We begin by writing down the metric form of the
constitutive relation that is known to reproduce the equivalent of General Relativity, then consider arbitrary constitutive relations in terms of a metric, and finally analyse in detail
the generic local and linear constitutive relations.

\subsection{Electromagnetism}
\label{electroconst}

In order to have a predictive theory, one has to specify the kinetic and the potential excitations. In the case of linear constitutive relation, the generic kinetic constitutive tensor density $\tilde{\chi}^{\mu\nu\alpha}=\tilde{\chi}^{[\mu\nu]\alpha}$ has
24 independent components and the generic potential constitutive tensor density $\tilde{\xi}^{\mu\alpha}$ has 16 components. The constitutive relations can be written as
\be \label{const}
H^{\mu\nu} = \tilde{\chi}^{\mu\nu\alpha}A_\alpha\,, \quad P^\mu = \tilde{\xi}^{\mu\alpha}A_\alpha\,.
\ee
In the case of the Proca theory, we have a metric $g_{\mu\nu}$ and its Levi-Civita covariant derivative $\mathcal{D}_\mu$ at hand, and can write the constitutive relations as $\tilde{\chi}^{\mu\nu\alpha} = 2\sqrt{-g}g^{\alpha[\nu}\mathcal{D}^{\mu]}$ and $\tilde{\xi}^{\mu\alpha}=m^2\sqrt{-g}g^{\mu\alpha}$.
If in addition to linearity, we require that the constitutive relation does not
involve derivative operators other than the field strength, the generic ansatz then reads
\be \label{ansatz}
H^{\mu\nu} = {\chi}^{\mu\nu\alpha\beta}F_{\alpha\beta} + {\chi}^{\mu\nu\alpha}A_\alpha\,, \quad P^\mu = {\xi}^{\mu\alpha\beta}F_{\alpha\beta} + {\xi}^{\mu\alpha}A_\alpha\,,
\ee
where $ {\chi}^{\mu\nu\alpha\beta}= {\chi}^{[\mu\nu]\alpha\beta}= {\chi}^{\mu\nu[\alpha\beta]}$ has 36 and ${\xi}^{\mu\alpha\beta}={\xi}^{\mu[\alpha\beta]}$ has 24 independent components.
In this formulation, the Proca theory is ${\chi}^{\mu\nu\alpha\beta}=\sqrt{-g}g^{\mu\alpha}g^{\nu\beta}$, ${\chi}^{\mu\nu\alpha}=0$, and
${\xi}^{\mu\alpha\beta}=0$, ${\xi}^{\mu\alpha}=m^2\sqrt{-g}g^{\mu\alpha}$. These constitutive relations have the exchange symmetry ${\chi}^{\mu\nu\alpha\beta} = {\chi}^{\alpha\beta\mu\nu}$,
${\xi}^{\mu\alpha}={\xi}^{\alpha\mu}$. If we consider constitutive relations built with a metric $g_{\mu\nu}$ and the Levi-Civita tensor density $\epsilon^{\alpha\beta\gamma\delta}$, it is impossible to have
nontrivial pieces ${\chi}^{\mu\nu\alpha}$ or ${\xi}^{\mu\alpha\beta}$ (as we shall soon see, the case is different with gravity, and in fact it is then crucial to take into account the piece corresponding to ${\chi}^{\mu\nu\alpha}$).

The general relation ${\chi}^{\mu\nu\alpha\beta}$ in (\ref{ansatz}) has been well studied, see \cite{hehl2003foundations,Rubilar:2007qm}, and its complete classification has been performed. In particular, the relation can be decomposed into the principal part (20 components), the skewon part (15 components) and the axion part (1 component). In a general theory, the propagating fields may feature, in addition to the familiar electromagnetic field, an axion, a dilaton and the more exotic skewon \cite{Hehl:2005xu,Ni:2014qfa}.

\subsection{Gravity}

In analogy to (\ref{const}), the gravitational constitution relations are now specified by the 960 independent components of the kinetic constitutive tensor density
$\chi^{\mu\nu}{}_{\alpha\rho\sigma}{}^\beta= \chi^{[\mu\nu]}{}_{\alpha\rho\sigma}{}^\beta = \chi^{\mu\nu}{}_{\alpha(\rho\sigma)}{}^\beta$ and the 1600 independent components of the potential constitutive tensor density $\xi^\alpha{}_{\mu\nu}{}^\beta{}_{\rho\sigma}= \xi^\alpha{}_{(\mu\nu)}{}^\beta{}_{\rho\sigma}
= \xi^\alpha{}_{\mu\nu}{}^\beta{}_{(\rho\sigma)}$ as
\be \label{gravconst}
H^{\mu\nu}{}_\alpha = \chi^{\mu\nu}{}_{\alpha\rho\sigma}{}^\beta Q_\beta{}^{\rho\sigma}\,, \quad P^\alpha{}_{\mu\nu} = \xi^\alpha{}_{\mu\nu}{}^\beta{}_{\rho\sigma}Q_\beta{}^{\rho\sigma}\,.
\ee
There are three symmetrisations we have imposed here in order to extract only the symmetric part of the gauge field $A^{\alpha\beta}{}_\mu$, such that we can identify
identify $\varphi^{(\alpha\beta)} = g^{\alpha\beta}$.
Since this $g^{\mu\nu}$ is the only tensor we have at hand in addition to the ones appearing in (\ref{gravconst}), the most economical way to proceed is to assume that the constitutive relation features only this tensor.
We emphasise that the object $\varphi^{\mu\nu}$ emerged in the construction of the theory as we noticed that the potential $A^{\mu\nu}{}_\alpha$ can always be reduced to $A_\alpha{}^{\mu\nu}=-\nabla_\alpha \varphi^{\mu\nu}$, so we have only changed the name of $\varphi^{(\mu\nu)}$ into $g^{\mu\nu}$ and of $\nabla_\alpha \varphi^{(\mu\nu)}$ into
$-Q_\alpha{}^{\mu\nu}$, and not added these tensors ad hoc.

Since our starting point was the conservation of energy and momentum, another natural assumption is that the theory should be invariant under translations, $g_{\mu\nu} \rightarrow g_{\mu\nu} + \mathcal{D}_{(\mu}X_{\nu)}$. It has been shown that these two requirements uniquely specify the theory, that in the gauge wherein the connection vanishes was called the Coincident General Relativity (CGR) \cite{BeltranJimenez:2017tkd}.
The constitutive relations for that theory are
\begin{subequations}
\label{cgrconst}
\ba
\label{cgrconst1}
 \prescript{}{\text{CGR}}\chi^{\mu\nu}{}_{\alpha\rho\sigma}{}^\beta & = & \sqrt{-g}M^2 \lp g_{\alpha(\rho}\delta_{\sigma)}^{[\mu}g^{\nu]\beta} - g_{\rho\sigma}\delta_{\alpha}^{[\mu}g^{\nu]\beta}+\delta^{[\mu}_\alpha\delta^{\nu]}_{(\rho}\delta^\beta_{\sigma)}\rp\,, \\ \label{cgrconst2}
 \prescript{}{\text{CGR}}\xi^\alpha{}_{\mu\nu}{}^\beta{}_{\rho\sigma} & = & - \frac{1}{4}\sqrt{-g} M^2\lp g^{\alpha\beta}g_{\mu(\rho}g_{\sigma)\nu}  - 2\delta^\beta_{(\nu}g_{\mu)(\sigma}\delta^\alpha_{\rho)}
 - g^{\alpha\beta}g_{\mu\nu}g_{\sigma\rho} + \delta^\alpha_{(\sigma}\delta^\beta_{\rho)}g_{\mu\nu} + \delta^\alpha_{(\mu}\delta^\beta_{\nu)}g_{\sigma\rho}\rp\,.
\ea
\end{subequations}
The mass scale $M$ gives the gravitational coupling, and is related to the Newton's constant $G$ scale by $1/M^2=8\pi G$.
There is a remarkable relation
\be
\text{in CGR}: \quad \nabla_\mu H^{\mu\nu}{}_\alpha = 2\nabla_\mu P^{\mu\nu}{}_\alpha\,. \quad
\ee
This relation is an identity that holds despite the different symmetries of the two tensor densities. Note that it implies that
\be
\text{in CGR}: \quad \nabla_\mu\nabla_\nu  P^{\mu\nu}{}_\alpha =0\,,
\ee
which indeed can be derived as a Bianchi identity  \cite{BeltranJimenez:2018vdo} .
Since in the equations of motion the tensor density $H^{\mu\nu}{}_\alpha$ only features in (\ref{gravity}), where it can be equivalently replaced
by (twice) the  tensor density $P^{\mu\nu}{}_\alpha$, it appears that at the level of dynamics, we can identify the latter as both the kinetic and the potential excitation tensor density.
Defining $\tau^\mu{}_\nu = \nabla_\alpha H^{\alpha\mu}{}_\nu$, we can rewrite the identity (\ref{gravity}) as
\be
\tau^\mu{}_\nu= T^\mu{}_\nu + t^\mu{}_\nu\,,
\ee
which features explicitly the canonical decomposition \cite{Jimenez:2019yyx} of the field equation into the gravitational, matter, and inertial energy-momentum tensor densities.
There is an important subtlety however, that since $\nabla_\sigma \varphi^{[\sigma\mu]\nu}{}_\alpha =  H^{\mu\nu}{}_\alpha - 2P^{\mu\nu}{}_\alpha \neq 0$, we should not employ the potential excitation to deduce the energy-momentum
of a gravitating system, but should use kinetic excitation.

\subsection{General metric constitutive relation}

Let us then continue to consider more general possibilities than the special case of CGR. The generic metric relation (\ref{const}) is given by 9 parameters as
\ba
\prescript{}{\text{EVEN}}\chi^{\mu\nu}{}_{\alpha\rho\sigma}{}^\beta & = & \sqrt{-g}\lp b_1 g_{\alpha(\rho}\delta_{\sigma)}^{[\mu}g^{\nu]\beta} + b_2 g_{\rho\sigma}\delta_{\alpha}^{[\mu}g^{\nu]\beta} + b_3 \delta^{[\mu}_\alpha\delta^{\nu]}_{(\rho}\delta^\beta_{\sigma)} \rp\,, \label{eq:chinewer}\\
 \prescript{}{\text{EVEN}}\xi^\alpha{}_{\mu\nu}{}^\beta{}_{\rho\sigma} & = & \sqrt{-g}\lp c_1 g^{\alpha\beta}g_{\mu(\rho}g_{\sigma)\nu}  + c_2 \delta^\beta_{(\nu}g_{\mu)(\sigma}\delta^\alpha_{\rho)}
 + c_3 g^{\alpha\beta}g_{\mu\nu}g_{\sigma\rho} + c_4\delta^\alpha_{(\mu}g_{\nu)(\rho}\delta^\beta_{\sigma)} +
 \frac{1}{2}\tilde{c}_5 \delta^\alpha_{(\sigma}\delta^\beta_{\rho)}g_{\mu\nu} + \frac{1}{2}\tilde{c}_6\delta^\alpha_{(\mu}\delta^\beta_{\nu)}g_{\sigma\rho}\rp\,. \quad \label{eq:xinewer}
\ea
For the two last pieces, it will be more convenient later to employ the different parameterisation
\be
\frac{1}{2}\tilde{c}_5 \delta^\alpha_{(\sigma}\delta^\beta_{\rho)}g_{\mu\nu} + \frac{1}{2}\tilde{c}_6\delta^\alpha_{(\mu}\delta^\beta_{\nu)}g_{\sigma\rho} =
\frac{1}{2}c_5\lp\delta^\alpha_{(\sigma}\delta^\beta_{\rho)}g_{\mu\nu} + \delta^\alpha_{(\mu}\delta^\beta_{\nu)}g_{\sigma\rho}\rp + \frac{1}{2}c_6\lp\delta^\alpha_{(\sigma}\delta^\beta_{\rho)}g_{\mu\nu} - \delta^\alpha_{(\mu}\delta^\beta_{\nu)}g_{\sigma\rho}\rp\,.
\ee
It can be seen that the $c_5$-term is reversible and $c_6$ is the coefficient of a skewon term.
As noticed by Vilson and R\"unkla \cite{Runkla:2018xrv}, the relation has the exchange symmetry $\xi^\alpha{}_{\mu\nu}{}^\beta{}_{\rho\sigma} = \xi^\beta{}_{\rho\sigma}{}^\alpha{}_{\mu\nu}$
when $\tilde{c}_6=\tilde{c}_5$, i.e. $c_6=0$. Only the components of the constitutive relation which have this symmetry enter into the Lagrangian, and having a Lagrangian formulation is a necessary condition for reversibility \cite{Itin:2018dru}.

The constitutive relation (\ref{eq:xinewer}) with $c_6=0$ describes the 5-parameter action of what was dubbed the Newer General Relativity \cite{BeltranJimenez:2017tkd} and has been studied
on many occasions \cite{Adak:2005cd,Adak:2008gd,BeltranJimenez:2018vdo,Runkla:2018xrv,Hohmann:2018wxu,Soudi:2018dhv}.
However, it is worth reiterating that the unique constitutive relation (\ref{cgrconst}) is dictated by $\nabla_\mu\nabla_\nu  P^{\mu\nu}{}_\alpha =0$ and $\nabla_\mu H^{\mu\nu}{}_\alpha = 2\nabla_\mu P^{\mu\nu}{}_\alpha$, which reflects the translational invariance of the purified gravity theory \cite{BeltranJimenez:2018vdo}. Conroy analysed the case of a generic linear constitutive relation without the restriction to first derivative order or even the assumption of locality \cite{Conroy:2017yln}, which required the parameterisation by 9 independent functions (the argument of those functions being the d'Alembertian operator $g^{\mu\nu}\nabla_\mu\nabla_\nu$).  Nonlinear constitutive relations have been applied in the context of $f(Q)$ cosmology\footnote{Such models may have also relevance at galactic scales \cite{Milgrom:2019rtd}.} \cite{BeltranJimenez:2017tkd,Harko:2018gxr,Lu:2019hra,Jimenez:2019ovq,Lazkoz:2019sjl,Xu:2019sbp}, and Dialektopoulos has classified the cosmological Noether symmetries of the most generic nonlinear first-derivative action \cite{Dialektopoulos:2019mtr}. The alternative possibilities that are uncovered in the premetric formalism could also be interesting to study in more detail.

In the case that one resorts, in addition to the metric, to the Levi-Civita tensor density $\epsilon^{\alpha\beta\gamma\delta}$, it is possible to consider parity-violating purified gravity theories \cite{Iosifidis:2018zwo,Conroy:2019ibo}.
Only one additional term can appear in the quadratic form, but four contractions can be formed in the kinetic constitutive relation. The parity-violating constitutive
relations are
\ba
 \prescript{}{\text{ODD}}\chi^{\mu\nu}{}_{\alpha\rho\sigma}{}^\beta & = & b_4\epsilon^{\mu\nu}{}_{\alpha(\rho}\delta^\beta_{\sigma)} +
 b_5\epsilon^{\mu\nu}{}_\alpha{}^\beta g_{\rho\sigma} + b_6\epsilon_\alpha{}^{\beta[\mu}{}_{(\rho}\delta^{\nu]}_{\sigma)} + b_7\epsilon^{\mu\nu\beta}{}_{(\rho}g_{\sigma)\alpha}\,, \label{eq:chiodd}\\
 \prescript{}{\text{ODD}}\xi^\alpha{}_{\mu\nu}{}^\beta{}_{\rho\sigma} & = & c_7\lp \epsilon^{\alpha\beta}{}_{\mu(\rho}g_{\sigma)\nu} + \epsilon^{\alpha\beta}{}_{\nu(\rho}g_{\sigma)\mu}\rp\,. \label{eq:xiodd}
\ea
The excitation tensor densities implied by the most general metric constitutive relation are therefore, explicitly,
\ba
H^{\mu\nu}{}_\alpha & = & -\sqrt{-g}\lp b_1 Q^{[\mu\nu]}{}_\alpha + b_2Q^{[\mu}\delta^{\nu]}_\alpha + b_ 3 \tilde{Q}^{[\mu}\delta^{\nu]}_\alpha\rp + b_4\epsilon^{\mu\nu}{}_{\alpha\beta}\tilde{Q}^\beta
+ b_5\epsilon^{\mu\nu}{}_{\alpha\beta}{Q}^\beta + b_6\epsilon_{\alpha}{}^{\beta[\mu}{}_\rho Q_\beta{}^{\nu]\rho}\,, \\
P^\alpha{}_{\mu\nu} & = & \sqrt{-g}\lb c_1 Q^\alpha{}_{\mu\nu} + c_2 Q_{(\mu}{}^\alpha{}_{\nu)} + c_3 Q^\alpha g_{\mu\nu} + c_4\delta^\alpha_{(\mu}\tilde{Q}_{\nu)}
+ \frac{1}{2}\lp c_5 \delta^\alpha_{(\mu}\tilde{Q}_{\nu)} + c_6\delta^\alpha_{(\mu}Q_{\nu)}\rp \rb - 2c_7\epsilon^{\alpha}{}_{\beta\rho(\mu}Q^{\beta}{}_{\nu)}{}^\rho\,.
\ea
However, the parity-violating constitution relation cannot satisfy the requirement $\nabla_\alpha H^{\alpha\mu}{}_\nu = 2\nabla_\alpha P^{\alpha\mu}{}_\nu$. The CGR relations
(\ref{cgrconst}) are therefore the unique local and linear (non-derivative) constitutive law of a Lagrangian theory, and this excludes odd-parity interactions.

Finally, we should remark that though the above requirement ensures that the premetric equations can be derived from an action principle, such a principle may not be necessary for a consistent
physical theory. If this assumption is relaxed, one may consider the full
13-component set of theories described by the above constitutive relations $\chi^{\mu\nu}{}_{\alpha\rho\sigma}{}^\beta= \prescript{}{\text{ODD}}\chi^{\mu\nu}{}_{\alpha\rho\sigma}{}^\beta + \prescript{}{\text{EVEN}}\chi^{\mu\nu}{}_{\alpha\rho\sigma}{}^\beta$ and $\xi^\alpha{}_{\mu\nu}{}^\beta{}_{\rho\sigma}= \prescript{}{\text{ODD}}\xi^\alpha{}_{\mu\nu}{}^\beta{}_{\rho\sigma} + \prescript{}{\text{EVEN}}\xi^\alpha{}_{\mu\nu}{}^\beta{}_{\rho\sigma}$. However, except
in the case of CGR, these theories may not be compatible with the metric-covariant conservation of matter energy-momentum (though they, by construction, are compatible with the
conservation of the total energy-momentum of the matter and the metric field) unless additional constraints are imposed.
As we will demonstrate in Section \ref{properties}, devoted to the further study of the possible viability of such generalised class of purified gravity (which could be dubbed the
Premetric Newer General Relativity), this turns out to be a case. Now we instead continue with the investigation of the constitutive relations, proceeding to the most generic case wherein no metric (nor Levi-Civita tensor) need be assumed.

\subsection{Irreducible decomposition of $\xi$}

Since the constitutive tensor densities $\xi^{\alpha}{}_{\mu\nu}{}^{\beta}{}_{\rho\sigma}$ and $\chi^{\mu\nu}{}_{\alpha\rho\sigma}{}^{\beta}$ are rather cumbersome objects with a large number of components, it is helpful to decompose them into smaller parts. Here we make use of an irreducible decomposition based on Young diagrams, along the lines of a similar decomposition shown in~\cite{Itin:2018dru}. We begin with the potential constitutive tensor density $\xi^{\alpha}{}_{\mu\nu}{}^{\beta}{}_{\rho\sigma} = \xi^{\alpha}{}_{(\mu\nu)}{}^{\beta}{}_{\rho\sigma} = \xi^{\alpha}{}_{\mu\nu}{}^{\beta}{}_{(\rho\sigma)}$. Its decomposition can be visualized in terms of Young diagrams as
\begin{equation}
\begin{split}
\yng(1) \otimes \yng(2) \otimes \yng(1) \otimes \yng(2) &= \left(\yng(1) \otimes \yng(1)\right) \otimes \left(\yng(2) \otimes \yng(2)\right)\\
&= \left(\yng(2) \oplus \yng(1,1)\right) \otimes \left(\yng(2,2) \oplus \yng(3,1) \oplus \yng(4)\right)\,,
\end{split}
\end{equation}
where the first bracket corresponds to the upper indices, while the second bracket corresponds to the lower indices. In 4 dimensions one finds that the total number of components decomposes as
\be
1600 = 200 \oplus 120 \oplus 450 \oplus 270 \oplus 350 \oplus 210\,.
\ee
By applying the Young projectors to $\xi^{\alpha}{}_{\mu\nu}{}^{\beta}{}_{\rho\sigma}$ one finds that its irreducible parts are given by
\begin{subequations}\label{eq:xidecomp}
\ba
\prescript{[1]}{}\xi^{\alpha}{}_{\mu\nu}{}^{\beta}{}_{\rho\sigma} & = & \xi^{(\alpha}{}_{\mu\nu}{}^{\beta)}{}_{\rho\sigma} - \prescript{[3]}{}\xi^{\alpha}{}_{\mu\nu}{}^{\beta}{}_{\rho\sigma} - \prescript{[5]}{}\xi^{\alpha}{}_{\mu\nu}{}^{\beta}{}_{\rho\sigma}\,,\\
\prescript{[2]}{}\xi^{\alpha}{}_{\mu\nu}{}^{\beta}{}_{\rho\sigma} & = & \xi^{[\alpha}{}_{\mu\nu}{}^{\beta]}{}_{\rho\sigma} - \prescript{[4]}{}\xi^{\alpha}{}_{\mu\nu}{}^{\beta}{}_{\rho\sigma} - \prescript{[6]}{}\xi^{\alpha}{}_{\mu\nu}{}^{\beta}{}_{\rho\sigma}\,,\\
\prescript{[3]}{}\xi^{\alpha}{}_{\mu\nu}{}^{\beta}{}_{\rho\sigma} & = & \frac{1}{2}\lp\xi^{(\alpha}{}_{\mu\nu}{}^{\beta)}{}_{\rho\sigma} - \xi^{(\alpha}{}_{\rho\sigma}{}^{\beta)}{}_{\mu\nu}\rp\,,\\
\prescript{[4]}{}\xi^{\alpha}{}_{\mu\nu}{}^{\beta}{}_{\rho\sigma} & = & \frac{1}{2}\lp\xi^{[\alpha}{}_{\mu\nu}{}^{\beta]}{}_{\rho\sigma} - \xi^{[\alpha}{}_{\rho\sigma}{}^{\beta]}{}_{\mu\nu}\rp\,,\\
\prescript{[5]}{}\xi^{\alpha}{}_{\mu\nu}{}^{\beta}{}_{\rho\sigma} & = & \xi^{(\alpha}{}_{(\mu\nu}{}^{\beta)}{}_{\rho\sigma)}\,,\\
\prescript{[6]}{}\xi^{\alpha}{}_{\mu\nu}{}^{\beta}{}_{\rho\sigma} & = & \xi^{[\alpha}{}_{(\mu\nu}{}^{\beta]}{}_{\rho\sigma)}\,.
\ea
\end{subequations}
The structure of this decomposition becomes clearer if we first decompose $\xi^{\alpha}{}_{\mu\nu}{}^{\beta}{}_{\rho\sigma}$ into its reversible and irreversible parts,
\begin{subequations}
\ba
\accentset{+}{\xi}^{\alpha}{}_{\mu\nu}{}^{\beta}{}_{\rho\sigma} & = & \frac{1}{2}\lp\xi^{\alpha}{}_{\mu\nu}{}^{\beta}{}_{\rho\sigma} + \xi^{\beta}{}_{\rho\sigma}{}^{\alpha}{}_{\mu\nu}\rp = \accentset{+}{\xi}^{\beta}{}_{\rho\sigma}{}^{\alpha}{}_{\mu\nu}\,,\\
\accentset{-}{\xi}^{\alpha}{}_{\mu\nu}{}^{\beta}{}_{\rho\sigma} & = & \frac{1}{2}\lp\xi^{\alpha}{}_{\mu\nu}{}^{\beta}{}_{\rho\sigma} - \xi^{\beta}{}_{\rho\sigma}{}^{\alpha}{}_{\mu\nu}\rp = -\accentset{-}{\xi}^{\beta}{}_{\rho\sigma}{}^{\alpha}{}_{\mu\nu}\,.
\ea
\end{subequations}
Note that only $\accentset{+}{\xi}^{\alpha}{}_{\mu\nu}{}^{\beta}{}_{\rho\sigma}$ contributes to the Lagrangian and preserves matter energy-momentum, while $\accentset{-}{\xi}^{\alpha}{}_{\mu\nu}{}^{\beta}{}_{\rho\sigma}$ mediates dissipative effects. We then further decompose these two parts by imposing the symmetry or antisymmetry of the upper two indices,
\be
\accentset{\pm}{\xi}^{\alpha}{}_{\mu\nu}{}^{\beta}{}_{\rho\sigma} = \accentset{\pm}{\xi}^{(\alpha}{}_{\mu\nu}{}^{\beta)}{}_{\rho\sigma} + \accentset{\pm}{\xi}^{[\alpha}{}_{\mu\nu}{}^{\beta]}{}_{\rho\sigma}\,.
\ee
Carefully examining the decomposition~\eqref{eq:xidecomp} then shows that it can alternatively be written in the equivalent form
\begin{subequations}\label{eq:xidecompri}
\ba
\prescript{[1]}{}\xi^{\alpha}{}_{\mu\nu}{}^{\beta}{}_{\rho\sigma} & = & \accentset{+}{\xi}^{(\alpha}{}_{\mu\nu}{}^{\beta)}{}_{\rho\sigma} - \accentset{+}{\xi}^{(\alpha}{}_{(\mu\nu}{}^{\beta)}{}_{\rho\sigma)}\,,\\
\prescript{[2]}{}\xi^{\alpha}{}_{\mu\nu}{}^{\beta}{}_{\rho\sigma} & = & \accentset{-}{\xi}^{[\alpha}{}_{\mu\nu}{}^{\beta]}{}_{\rho\sigma} - \accentset{-}{\xi}^{[\alpha}{}_{(\mu\nu}{}^{\beta]}{}_{\rho\sigma)}\,,\\
\prescript{[3]}{}\xi^{\alpha}{}_{\mu\nu}{}^{\beta}{}_{\rho\sigma} & = & \accentset{-}{\xi}^{[\alpha}{}_{\mu\nu}{}^{\beta]}{}_{\rho\sigma}\,,\\
\prescript{[4]}{}\xi^{\alpha}{}_{\mu\nu}{}^{\beta}{}_{\rho\sigma} & = & \accentset{+}{\xi}^{(\alpha}{}_{\mu\nu}{}^{\beta)}{}_{\rho\sigma}\,,\\
\prescript{[5]}{}\xi^{\alpha}{}_{\mu\nu}{}^{\beta}{}_{\rho\sigma} & = & \accentset{+}{\xi}^{(\alpha}{}_{(\mu\nu}{}^{\beta)}{}_{\rho\sigma)}\,,\\
\prescript{[6]}{}\xi^{\alpha}{}_{\mu\nu}{}^{\beta}{}_{\rho\sigma} & = & \accentset{-}{\xi}^{[\alpha}{}_{(\mu\nu}{}^{\beta]}{}_{\rho\sigma)}\,.
\ea
\end{subequations}
The reversible and irreversible parts thus decompose as
\begin{subequations}
\ba
\accentset{+}{\xi}^{\alpha}{}_{\mu\nu}{}^{\beta}{}_{\rho\sigma} & = & \prescript{[1]}{}\xi^{\alpha}{}_{\mu\nu}{}^{\beta}{}_{\rho\sigma} + \prescript{[4]}{}\xi^{\alpha}{}_{\mu\nu}{}^{\beta}{}_{\rho\sigma} + \prescript{[5]}{}\xi^{\alpha}{}_{\mu\nu}{}^{\beta}{}_{\rho\sigma}\,,\\
\accentset{-}{\xi}^{\alpha}{}_{\mu\nu}{}^{\beta}{}_{\rho\sigma} & = & \prescript{[2]}{}\xi^{\alpha}{}_{\mu\nu}{}^{\beta}{}_{\rho\sigma} + \prescript{[3]}{}\xi^{\alpha}{}_{\mu\nu}{}^{\beta}{}_{\rho\sigma} + \prescript{[6]}{}\xi^{\alpha}{}_{\mu\nu}{}^{\beta}{}_{\rho\sigma}\,.
\ea
\end{subequations}
In Table \ref{nomenclature} we give names to these irreducible pieces, following a logic adapted from~\cite{Itin:2018dru}. All parts contribute to energy-momentum, but not all can be derived from a Lagrangian formulation.

\begin{center}
\begin{table}[h]
\begin{tabular}{|c| c| c| c| c|| c |}
Irreducible part & Components & Nomenclature & Lagrangian & Dispersion & Metric terms  \\
\hline
$\prescript{[1]}{}\xi^{\alpha}{}_{\mu\nu}{}^{\beta}{}_{\rho\sigma}$ &  200  & Principal-1 & Yes & - & $c_1-2c_3$, $c_2+c_4-2c_5$ \\
$\prescript{[2]}{}\xi^{\alpha}{}_{\mu\nu}{}^{\beta}{}_{\rho\sigma}$ &  120 & Skewon-1 & No  & - & none \\
$\prescript{[3]}{}\xi^{\alpha}{}_{\mu\nu}{}^{\beta}{}_{\rho\sigma}$ &  450 & Skewon-2 & No  & - & $c_6$ \\
$\prescript{[4]}{}\xi^{\alpha}{}_{\mu\nu}{}^{\beta}{}_{\rho\sigma}$ &  270 & Axion-1 & Yes  & - & $c_4-c_2$, $c_7$ \\
$\prescript{[5]}{}\xi^{\alpha}{}_{\mu\nu}{}^{\beta}{}_{\rho\sigma}$ &  350 & Principal-2 & Yes  & - & $c_1+c_3$, $c_2+c_4+c_5$ \\
$\prescript{[6]}{}\xi^{\alpha}{}_{\mu\nu}{}^{\beta}{}_{\rho\sigma}$ &  210 & Axion-2 & No & - &  none \\
\hline
$\prescript{[1]}{}\chi^{\mu\nu}{}_{\alpha\rho\sigma}{}^{\beta}$ & 400 & Principal-A & (Yes) & Yes & $b_1-2b_2$, $2b_4-b_6$  \\
$\prescript{[2]}{}\chi^{\mu\nu}{}_{\alpha\rho\sigma}{}^{\beta}$ &  400 & Principal-B & (Yes) & Yes & $b_1+b_2$ \\
$\prescript{[3]}{}\chi^{\mu\nu}{}_{\alpha\rho\sigma}{}^{\beta}$ &  80 & Axion-A & (No) & No & $b_4+b_6$, $2b_5+b_7$ \\
$\prescript{[4]}{}\chi^{\mu\nu}{}_{\alpha\rho\sigma}{}^{\beta}$ &  80 & Axion-B & (No) & No & $b_5-b_7$ \\
\hline
$\prescript{\{1\}}{}\chi^{\mu\nu}{}_{\alpha\rho\sigma}{}^{\beta}$ & 336 & Odd Axion & (No) & (No) & $b_5-b_7$ \\
$\prescript{\{2\}}{}\chi^{\mu\nu}{}_{\alpha\rho\sigma}{}^{\beta}$ &  240 & Odd Principal-1 & (No) & (Yes) & $2b_5+b_7-b_4$, $2b_5+b_7$ \\
$\prescript{\{3\}}{}\chi^{\mu\nu}{}_{\alpha\rho\sigma}{}^{\beta}$ &  2 $\cdot$ 144 & Even Principal-1 & (Yes) & (Yes) & $b_1$, $b_2$ \\
$\prescript{\{4\}}{}\chi^{\mu\nu}{}_{\alpha\rho\sigma}{}^{\beta}$ &  80 & Odd Principal-2 & (No) & (Yes) & $b_4-2b_5-2b_6-b_7$ \\
$\prescript{\{5\}}{}\chi^{\mu\nu}{}_{\alpha\rho\sigma}{}^{\beta}$ &  16 & Even Principal-2 & (Yes) & (Yes) & $b_1-2b_2-2b_3$ \\
\hline
\end{tabular}
\caption{Nomenclature for the irreducible parts of the constitutive relations. In the last column we indicate which combinations of the metric terms contribute to each of the irreducible parts.
Elsewhere, the parenthesis is used to indicate that we have a definite answer only in a metrical theory.
As will be clarified later, the dispersion relation of gravitational waves depends at the linear order only on the constitutive relation $\chi^{\mu\nu}{}_{\alpha\rho\sigma}{}^{\beta}$,
and not all its irreducible parts contribute.
\label{nomenclature}}
\end{table}
\end{center}

\subsection{Irreducible decomposition of $\chi$}

We then continue with the kinetic constitutive tensor density $\chi^{\mu\nu}{}_{\alpha\rho\sigma}{}^\beta = \chi^{[\mu\nu]}{}_{\alpha\rho\sigma}{}^\beta = \chi^{\mu\nu}{}_{\alpha(\rho\sigma)}{}^\beta$. In terms of Young diagrams the decomposition is given by
\begin{equation}
\begin{split}
\yng(1,1) \otimes \yng(1) \otimes \yng(2) \otimes \yng(1) &= \left(\yng(1,1) \otimes \yng(1)\right) \otimes \left(\yng(1) \otimes \yng(2)\right)\\
&= \left(\yng(2,1) \oplus \yng(1,1,1)\right) \otimes \left(\yng(2,1) \oplus \yng(3)\right)\,,
\end{split}
\end{equation}
where again the first and second bracket correspond to upper and lower indices, respectively. One then finds that the total number of independent components splits {into}
\be
960 = 400 \oplus 400 \oplus 80 \oplus 80\,.
\ee
The irreducible decomposition of $\chi^{\mu\nu}{}_{\alpha\rho\sigma}{}^{\beta}$ is given by
\begin{subequations}\label{eq:chidecomp}
\ba
\prescript{[1]}{}\chi^{\mu\nu}{}_{\alpha\rho\sigma}{}^{\beta} & = & \chi^{\mu\nu}{}_{\alpha\rho\sigma}{}^{\beta} - \prescript{[2]}{}\chi^{\mu\nu}{}_{\alpha\rho\sigma}{}^{\beta} - \prescript{[3]}{}\chi^{\mu\nu}{}_{\alpha\rho\sigma}{}^{\beta} - \prescript{[4]}{}\chi^{\mu\nu}{}_{\alpha\rho\sigma}{}^{\beta}\,,\\
\prescript{[2]}{}\chi^{\mu\nu}{}_{\alpha\rho\sigma}{}^{\beta} & = & \chi^{\mu\nu}{}_{(\alpha\rho\sigma)}{}^{\beta} - \prescript{[4]}{}\chi^{\mu\nu}{}_{\alpha\rho\sigma}{}^{\beta}\,,\\
\prescript{[3]}{}\chi^{\mu\nu}{}_{\alpha\rho\sigma}{}^{\beta} & = & \chi^{[\mu\nu}{}_{\alpha\rho\sigma}{}^{\beta]} - \prescript{[4]}{}\chi^{\mu\nu}{}_{\alpha\rho\sigma}{}^{\beta}\,,\\
\prescript{[4]}{}\chi^{\mu\nu}{}_{\alpha\rho\sigma}{}^{\beta} & = & \chi^{[\mu\nu}{}_{(\alpha\rho\sigma)}{}^{\beta]}\,.
\ea
\end{subequations}
An alternative decomposition can be obtained by lowering the first pair of indices {by} using the Levi-Civita symbol, and defining
\begin{equation}
\tilde{\chi}_{\mu\nu\alpha\rho\sigma}{}^{\beta} = \frac{1}{2}\epsilon_{\mu\nu\tau\omega}\chi^{\tau\omega}{}_{\alpha\rho\sigma}{}^{\beta}\,.
\end{equation}
We can then perform a decomposition in the lower indices, which is expressed in Young diagrams as
\begin{equation}
\yng(1,1) \otimes \yng(1) \otimes \yng(2) = \yng(4,1) \oplus \yng(3,2) \oplus 2 \cdot \yng(3,1,1) \oplus \yng(2,2,1) \oplus \yng(2,1,1,1)\,.
\end{equation}
Taking into account also the upper index, which we omitted in the decomposition above, the number of independent components splits as
\begin{equation}
960 = 336 \oplus 240 \oplus 2 \cdot 144 \oplus 80 \oplus 16\,.
\end{equation}
Particular attention should be paid to the third diagram, which appears twice in the decomposition. This indicates that the irreducible tensor decomposition, seen as a decomposition of a tensor product of representations of \(\mathrm{GL}(4)\) into irreducible subrepresentations, contains two copies of the same irreducible representation represented by this diagram. However, in contrast to the remaining representations, which appear only once in the decomposition, there is no \emph{canonical} choice of the two representation spaces (and hence projectors onto particular tensor components); only their direct sum is canonically determined. Thus, the decomposition yields 5 terms, which we label \(\prescript{\{I\}}{}{\tilde{\chi}}_{\mu\nu\alpha\rho\sigma}{}^{\beta}, I = 1, \ldots, 5\). Keeping in mind that these are still antisymmetric in the first two indices, we may raise these indices again, hence defining
\begin{equation}
\prescript{\{I\}}{}\chi^{\mu\nu}{}_{\alpha\rho\sigma}{}^{\beta} = -\frac{1}{2}\epsilon^{\mu\nu\tau\omega}\prescript{\{I\}}{}{\tilde{\chi}}_{\tau\omega\alpha\rho\sigma}{}^{\beta}\,.
\end{equation}
These terms are then given by
\begin{subequations}\label{eq:chidecomp2}
\begin{align}
\prescript{\{1\}}{}\chi^{\mu\nu}{}_{\alpha\rho\sigma}{}^{\beta} &= \frac{1}{5}\left(\chi^{\mu\nu}{}_{(\alpha\rho\sigma)}{}^{\beta} + 2\delta_{(\alpha}^{[\mu}\chi^{\nu]\gamma}{}_{|\gamma|\rho\sigma)}{}^{\beta} + 4\delta_{(\alpha}^{[\mu}\chi^{\nu]\gamma}{}_{\rho\sigma)\gamma}{}^{\beta}\right)\,,\\
\prescript{\{2\}}{}\chi^{\mu\nu}{}_{\alpha\rho\sigma}{}^{\beta} &= \chi^{\mu\nu}{}_{\alpha\rho\sigma}{}^{\beta} - \chi^{\mu\nu}{}_{(\alpha\rho\sigma)}{}^{\beta} - \frac{1}{2}\chi^{\gamma\delta}{}_{\gamma\delta(\rho}{}^{\beta}\delta_{\sigma)}^{[\mu}\delta_{\alpha}^{\nu]} + \frac{2}{3}\delta_{\alpha}{}^{[\mu}\chi^{\nu]\gamma}{}_{\gamma\rho\sigma}{}^{\beta}\nonumber\\
&\phantom{=}+ \frac{5}{6}\delta_{(\rho}^{[\mu}\chi^{\nu]\gamma}{}_{|\alpha\gamma|\sigma)}{}^{\beta} - \frac{1}{6}\delta_{(\rho}^{[\mu}\chi^{\nu]\gamma}{}_{\sigma)\alpha\gamma}{}^{\beta} - \frac{2}{3}\delta_{(\rho}^{[\mu}\chi^{\nu]\gamma}{}_{|\gamma\alpha|\sigma)}{}^{\beta} - \frac{2}{3}\delta_{\alpha}{}^{[\mu}\chi^{\nu]\gamma}{}_{(\rho\sigma)\gamma}{}^{\beta}\,,\\
\prescript{\{3\}}{}\chi^{\mu\nu}{}_{\alpha\rho\sigma}{}^{\beta} &= \frac{1}{5}\Big(3\chi^{\gamma\delta}{}_{\gamma\delta(\rho}{}^{\beta}\delta_{\sigma)}^{[\mu}\delta_{\alpha}^{\nu]} - 3\delta_{(\rho}^{[\mu}\chi^{\nu]\gamma}{}_{\sigma)\alpha\gamma}{}^{\beta} - 3\delta_{(\rho}^{[\mu}\chi^{\nu]\gamma}{}_{|\alpha\gamma|\sigma)}{}^{\beta}\nonumber\\
&\phantom{=}+ 2\delta_{(\rho}^{[\mu}\chi^{\nu]\gamma}{}_{|\gamma\alpha|\sigma)}{}^{\beta} + 2\delta_{\alpha}{}^{[\mu}\chi^{\nu]\gamma}{}_{(\rho\sigma)\gamma}{}^{\beta} - 4\delta_{\alpha}{}^{[\mu}\chi^{\nu]\gamma}{}_{\gamma\rho\sigma}{}^{\beta}\Big)\,,\\
\prescript{\{4\}}{}\chi^{\mu\nu}{}_{\alpha\rho\sigma}{}^{\beta} &= \frac{1}{2}\delta_{(\rho}^{[\mu}\chi^{\nu]\gamma}{}_{\sigma)\gamma\alpha}{}^{\beta} - \frac{1}{2}\delta_{(\rho}^{[\mu}\chi^{\nu]\gamma}{}_{|\alpha\gamma|\sigma)}{}^{\beta} + \frac{1}{6}\chi^{\gamma\delta}{}_{\gamma\delta(\rho}{}^{\beta}\delta_{\sigma)}^{[\mu}\delta_{\alpha}^{\nu]}\,,\\
\prescript{\{5\}}{}\chi^{\mu\nu}{}_{\alpha\rho\sigma}{}^{\beta} &= -\frac{4}{15}\chi^{\gamma\delta}{}_{\gamma\delta(\rho}{}^{\beta}\delta_{\sigma)}^{[\mu}\delta_{\alpha}^{\nu]}\,.
\end{align}
\end{subequations}
Some properties of these components are summarised in Table \ref{nomenclature}.

\subsection{Irreducible decomposition of metric constitutive law}

We now apply the decompositions shown above to the metric constitutive relations. The potential constitutive tensor density~(\ref{eq:xinewer},\ref{eq:xiodd}) decomposes into the parts
(in this subsection, we shall absorb the scale $M^2$ into the coefficients $c_i$ and $b_i$, which then become dimensionful)
\begin{subequations}
\ba
\prescript{[1]}{}\xi^{\alpha}{}_{\mu\nu}{}^{\beta}{}_{\rho\sigma} & = & \sqrt{-g}\lb\frac{c_1 - 2c_3}{3}g^{\alpha\beta}\lp g_{\mu(\rho}g_{\sigma)\nu} - g_{\mu\nu}g_{\rho\sigma}\rp + \frac{c_2 + c_4 - 2c_5}{6}\lp 2\delta_{(\mu}^{(\alpha}g_{\nu)(\rho}\delta_{\sigma)}^{\beta)} - g_{\mu\nu}\delta_{(\rho}^{\alpha}\delta_{\sigma)}^{\beta} - g_{\rho\sigma}\delta_{(\mu}^{\alpha}\delta_{\nu)}^{\beta}\rp\rb\,,\\
\prescript{[2]}{}\xi^{\alpha}{}_{\mu\nu}{}^{\beta}{}_{\rho\sigma} & = & 0\,,\\
\prescript{[3]}{}\xi^{\alpha}{}_{\mu\nu}{}^{\beta}{}_{\rho\sigma} & = & \sqrt{-g}\frac{c_6}{2}\lp\delta^\alpha_{(\sigma}\delta^\beta_{\rho)}g_{\mu\nu} - \delta^\alpha_{(\mu}\delta^\beta_{\nu)}g_{\sigma\rho}\rp\,,\\
\prescript{[4]}{}\xi^{\alpha}{}_{\mu\nu}{}^{\beta}{}_{\rho\sigma} & = & \sqrt{-g}(c_4 - c_2)\delta_{(\mu}^{[\alpha}g_{\nu)(\rho}\delta_{\sigma)}^{\beta]} + c_7\lp \epsilon^{\alpha\beta}{}_{\mu(\rho}g_{\sigma)\nu} + \epsilon^{\alpha\beta}{}_{\nu(\rho}g_{\sigma)\mu}\rp\,,\\
\prescript{[5]}{}\xi^{\alpha}{}_{\mu\nu}{}^{\beta}{}_{\rho\sigma} & = & \sqrt{-g}\lb(c_1 + c_3)g^{\alpha\beta}g_{(\mu\nu}g_{\rho\sigma)} + (c_2 + c_4 + c_5)g_{(\mu\nu}\delta_{\rho}^{\alpha}\delta_{\sigma)}^{\beta}\rb\,,\\
\prescript{[6]}{}\xi^{\alpha}{}_{\mu\nu}{}^{\beta}{}_{\rho\sigma} & = & 0\,.
\ea
\end{subequations}
We find that the only irreversible part is the skewon $\prescript{[3]}{}\xi^{\alpha}{}_{\mu\nu}{}^{\beta}{}_{\rho\sigma}$, which is non-vanishing only in the case $c_6 \neq 0$. We also find that the parity-violating term~(\ref{eq:xiodd}) contributes only to the part $\prescript{[4]}{}\xi^{\alpha}{}_{\mu\nu}{}^{\beta}{}_{\rho\sigma}$. For the kinetic constitutive tensor density~(\ref{eq:chinewer},\ref{eq:chiodd}) we have the irreducible parts
\begin{subequations}
\ba
\prescript{[1]}{}\chi^{\mu\nu}{}_{\alpha\rho\sigma}{}^{\beta} & = & \sqrt{-g}\lb\frac{b_1 - 2b_2}{3}\lp g_{\alpha(\rho}\delta_{\sigma)}^{[\mu}g^{\nu]\beta} - g_{\rho\sigma}\delta_{\alpha}^{[\mu}g^{\nu]\beta}\rp + b_3\delta_{\alpha}^{[\mu}\delta_{(\rho}^{\nu]}\delta_{\sigma)}^{\beta}\rb + \frac{2b_4 - b_6}{3}\lp\epsilon_{\alpha}{}^{\mu\nu}{}_{(\rho}\delta_{\sigma)}^{\beta} - \epsilon_{\alpha}{}^{\beta[\mu}{}_{(\rho}\delta_{\sigma)}^{\nu]}\rp\,,\\
\prescript{[2]}{}\chi^{\mu\nu}{}_{\alpha\rho\sigma}{}^{\beta} & = & \sqrt{-g}(b_1 + b_2)g_{(\rho\sigma}\delta_{\alpha)}^{[\mu}g^{\nu]\beta}\,,\\
\prescript{[3]}{}\chi^{\mu\nu}{}_{\alpha\rho\sigma}{}^{\beta} & = & (b_4 + b_6)\epsilon_{\alpha(\rho}{}^{[\mu\nu}\delta_{\sigma)}^{\beta]} + \frac{2b_5 + b_7}{3}\lp g_{\rho\sigma}\epsilon_{\alpha}{}^{\beta\mu\nu} - g_{\alpha(\rho}\epsilon_{\sigma)}{}^{\beta\mu\nu}\rp\,,\\
\prescript{[4]}{}\chi^{\mu\nu}{}_{\alpha\rho\sigma}{}^{\beta} & = & (b_5 - b_7)g_{(\rho\sigma}\epsilon_{\alpha)}{}^{\beta\mu\nu}\,.
\ea
\end{subequations}
Alternatively, we may use the decomposition~(\ref{eq:chidecomp2}) and find
\begin{subequations}
\begin{align}
\prescript{\{1\}}{}\chi^{\mu\nu}{}_{\alpha\rho\sigma}{}^{\beta} &= (b_5 - b_7)g_{(\rho\sigma}\epsilon_{\alpha)}{}^{\beta\mu\nu}\,,\\
\prescript{\{2\}}{}\chi^{\mu\nu}{}_{\alpha\rho\sigma}{}^{\beta} &= \frac{2b_5 + b_7 - b_4}{2}\delta_{(\rho}^{[\mu}\epsilon^{\nu]}{}_{\sigma)\alpha}{}^{\beta} + b_4\epsilon^{\mu\nu}{}_{\alpha(\rho}\delta_{\sigma)}^{\beta} + \frac{2b_5 + b_7}{3}\left(g_{\rho\sigma}\epsilon^{\mu\nu}{}_{\alpha}{}^{\beta} - g_{\alpha(\rho}\epsilon^{\mu\nu}{}_{\sigma)}{}^{\beta}\right)\,,\\
\prescript{\{3\}}{}\chi^{\mu\nu}{}_{\alpha\rho\sigma}{}^{\beta} &= \sqrt{-g}\left(\frac{b_1 - 2b_2}{5}\delta_{\alpha}^{[\mu}\delta_{(\rho}^{\nu]}\delta_{\sigma)}^{\beta} + b_1g_{\alpha(\rho}\delta_{\sigma)}^{[\mu}g^{\nu]\beta} + b_2g_{\rho\sigma}\delta_{\alpha}^{[\mu}g^{\nu]\beta}\right)\,,\\
\prescript{\{4\}}{}\chi^{\mu\nu}{}_{\alpha\rho\sigma}{}^{\beta} &= \frac{b_4 - 2b_5 - 2b_6 - b_7}{2}\delta_{(\rho}^{[\mu}\epsilon^{\nu]}{}_{\sigma)\alpha}{}^{\beta}\,,\\
\prescript{\{5\}}{}\chi^{\mu\nu}{}_{\alpha\rho\sigma}{}^{\beta} &= -\sqrt{-g}\frac{b_1 - 2b_2 - 5b_3}{5}\delta_{\alpha}^{[\mu}\delta_{(\rho}^{\nu]}\delta_{\sigma)}^{\beta}\,.
\end{align}
\end{subequations}
Following this decomposition, we find that the parity-preserving terms contribute only to $\prescript{\{3\}}{}\chi^{\mu\nu}{}_{\alpha\rho\sigma}{}^{\beta}$ and $\prescript{\{5\}}{}\chi^{\mu\nu}{}_{\alpha\rho\sigma}{}^{\beta}$, while the parity-violating terms contribute only to $\prescript{\{1\}}{}\chi^{\mu\nu}{}_{\alpha\rho\sigma}{}^{\beta}$, $\prescript{\{2\}}{}\chi^{\mu\nu}{}_{\alpha\rho\sigma}{}^{\beta}$ and $\prescript{\{4\}}{}\chi^{\mu\nu}{}_{\alpha\rho\sigma}{}^{\beta}$.

We shall return in Section \ref{conclusions} to summarise in Figure \ref{fig2} the physical assumptions which lead from the 5632-component general constitutive relation to the unique theory specified by 1 free component.

\section{Premetric construction in the language of differential forms}
\label{forms}

In this Section we revisit the steps of Section \ref{tensors} in a different formalism. A dictionary between the two languages will be given in Table \ref{table1}.

\subsection{Frame}

Consider a 4-dimensional differential manifold endowed with a coframe $\e^a$. Its exterior products generate the bases
\be \label{bases}
\e^a\,, \quad \e^{ab} = \e^a\wedge\e^b\,, \quad \e^{abc} = \e^a\wedge\e^b\wedge\e^c\,, \quad
\e^{abcd} = \e^a\wedge\e^b\wedge\e^c\wedge\e^d\,,
\ee
of the spaces of untwisted 1-forms, 2-forms, 3-forms and 4-forms, respectively. In terms of the Levi-Civita permutation symbol $\varepsilon_{acbd}$ and a section $s$ of the orientation line bundle, we can introduce the twisted scalar-valued volume form
\be
\text{vol} = \frac{1}{4!}\varepsilon_{acbd}\e^{abcd}\otimes s\,.
\ee
We note that we can use the interior product (to be denoted here with a ``dot'' instead of the perhaps more common ``hook'') without invoking a spacetime metric, it being in basic terms just a summation than a contraction.
Thus we can introduce the frame field $\ie_a$ as the inverse of the coframe,
\be
\ie_a\cdot\e^b = \e^b\cdot\ie_a = \delta^b_a\,,
\ee
and this allows to also introduce the basis form for the spaces of twisted 0-forms, 1-forms, 2-forms, 3-forms and 4-forms as
\be \label{tbases}
\epsilon_{abcd} = \ie_d\cdot\epsilon_{abc}\,, \quad \epsilon_{abc} = \ie_c\cdot\epsilon_{ab}\,, \quad
 \epsilon_{ab} = \ie_b\cdot\epsilon_{a}\,, \quad \epsilon_a = \ie_a\cdot \text{vol}\,, \quad \text{vol}\,,
\ee
respectively. One may check that $\e^a\wedge\epsilon_b=\delta^a_b\text{vol}$. Under a GL(4) transformation $\Lambda_a{}^b$ (with the inverse $\Lambda^a{}_b$ and the determinant $\Lambda)$, we have the following transformation laws:
\be \label{trans}
\e^a \rightarrow \Lambda_b{}^a \e^b\,, \quad \ie_a \rightarrow \Lambda^b{}_a\ie_b\,, \quad \text{vol} \rightarrow |\Lambda | \text{vol}\,,
\ee
and thus the bases (\ref{bases}) are tensors whilst the twisted bases (\ref{tbases}) are tensor densities.

\subsection{Excitation}

\subsubsection{Electromagnetism}

The conservation of the electric charge entails the existence of an electric current $\bJ$. It is described as a twisted 3-form,
\be
\bJ = J^a \epsilon_a\,.
\ee
Under the transformation (\ref{trans}) $J^a \rightarrow \Lambda^{-1}\Lambda^a{}_b J^b$, and thus $\bJ \rightarrow \pm \bJ$, where the sign is the
sign of $\Lambda$. The charge conservation, in integral and in differential forms is
\be
\int_{\partial \Omega_4}\bJ =0\,, \quad \text{and} \quad \diff \bJ = 0\,,
\ee
respectively. Locally, the latter is equivalent to the inhomogeneous Maxwell equation
\be
\diff \bH = \bJ\,,
\ee
implying the existence of the electromagnetic excitation $\bH$, which is a twisted 2-form
\be
\bH = \frac{1}{2}H_{ab}\e^{ab} = \frac{1}{2}\tilde{H}^{ab}\epsilon_{ab}\,, \quad \text{where} \quad \tilde{H}^{ab}=\frac{1}{2}\epsilon^{abcd}H_{cd}\,.
\ee
Since $H_{ab}$ is a twisted covariant tensor, $\tilde{H}^{ab}$ is an untwisted contravariant tensor density. To take into account that in addition to the
matter sources $\bT$, in massive electromagnetism there exists also the self-interaction source $\bt$, one performs the decomposition $\bJ = \bT + \bt$.

In case of such self-interactions, one also needs to consider a twisted 3-form $\bP$, given as
\be
\bP = \frac{1}{6}P_{abc}\e^{abc} = \tilde{P}^{a}\epsilon_{a}\,, \quad \text{where} \quad \tilde{P}^{a}=\frac{1}{6}\epsilon^{abcd}P_{bcd}\,.
\ee
Since $P_{abc}$ is a twisted covariant rank-3 tensor, $\tilde{P}^{a}$ is an untwisted contravariant vector density.

\subsubsection{Gravity}

In gravity, we begin with the conservation of energy and momenta, and we have thus 4 conserved charges.
As in the case of electromagnetism, they are described by a twisted 3-form,
\be
\bJ_{a} = J_{a}{}^c\epsilon_c\,.
\ee
The conservation in integral and in differential forms is analogously expressed as
\be
\int_{\partial \Omega_4}\bJ_{a} =0\,, \quad \text{and} \quad \diff \bJ_{a} = 0\,.
\ee
The latter implies again the existence of a twisted two-form
\be
\bH_{a} = \frac{1}{2}H_{abc}\e^{bc} = \frac{1}{2}\tilde{H}_{a}{}^{bc}\epsilon_{bc}\,, \quad \text{where} \quad \tilde{H}_{a}{}^{bc}=\frac{1}{2}\epsilon^{bcde}H_{ade}\,.
\ee
Since $H_{abc}$ is a twisted covariant tensor, $\tilde{H}_{a}{}^{bc}$ is an untwisted contravariant tensor density. We now write
\be
\diff \bH_{a} = \bJ_{a} = \bT_{a} + \bt_{a}\,,
\ee
taking into account that in addition to the energy-momentum of matter $\bT_{a}$, there can also occur inertial energy-momentum $\bt_{a}$.

The potential excitation is now defined as the twisted 3-form $\bP_{ab}$, given as
\be
\bP_{ab} = \frac{1}{6}P_{abcde}\e^{cde} = \tilde{P}_{ab}{}^{c}\epsilon_{c}\,, \quad \text{where} \quad \tilde{P}_{ab}{}^{c}=\frac{1}{6}\epsilon^{cdef}P_{abdef}\,.
\ee
Since $P_{abcde}$ is a twisted covariant rank-3 tensor, $\tilde{P}_{ab}{}^{c}$ is an untwisted contravariant vector density.

\subsection{Field strength}

\subsubsection{Electromagnetism}

The field strength $\bF=\frac{1}{2}F_{ab}\e^{ab}$ is an untwisted 2-form, which satisfies the equations
\be
\int_{\partial \Omega_4}\bF =0\,, \quad \text{and} \quad \diff \bF = 0\,.
\ee
The latter equation is an expression of the conservation of the magnetic flux, and it implies the existence of the electromagnetic potential
$\bA = A_a\e^a$ such that $\diff\bA = \bF$. The electromagnetic potential is defined up to a scalar $\varphi$ such that $\bA \rightarrow \bA + \diff\varphi$.

\subsubsection{Gravity}

We introduce the gravitational field strength $\bF^{ab}$ as an untwisted tensor-valued 2-form which satisfies the equations
$\bF^{ab} = 0$, due to the integrability of the gravitational geometry. The gravitational potential $\bA^{ab}=\bA^{ab}{}_{c}\e^c$ for which
$\bF^{ab} = \diff \bA^{ab}$, thus has the further property that $\bA^{ab} = \diff \varphi^{ab}$ for some $\varphi^{ab}$, which follows from our basic postulate that $\bF^{ab} = 0$.

\subsection{Force}

\subsubsection{Electromagnetism}

The force acting on matter is described by a covector-valued twisted 4-form $\boldsymbol{f}_a = f_a \text{vol}$, where $f_a$ is a covector-valued scalar.
The Lorentz force is given as
\be
\boldsymbol{f}_a = \lp\ie_a\cdot \bF\rp\wedge \bJ = \lp J^b F_{ba}\rp\text{vol} = f_a \text{vol}\,.
\ee
By construction, $f_a$ is an untwisted covector-valued scalar density.

\subsubsection{Gravity}

The gravitational force is, again, constructed in complete analogy to the electromagnetic one, as a covector-valued twisted 4-form $\boldsymbol{f}_{a}{}^b = f_{a}{}^b\text{vol}$,
\be
\boldsymbol{f}_{a}{}^c = \lp\ie_a\cdot \bF^{bc}\rp\wedge \bJ_{b} = \lp J_{b}{}^d F^{bc}{}_{da}\rp\text{vol} = f_a{}^c \text{vol}\,.
\ee
The absence of curvature, $\bF^{ab}=0$, implies the absence of force, $\boldsymbol{f}_{a}{}^b =0$. This reflects the conservation of total energy and momentum.

However, matter energy-momentum need not be conserved. Therefore, there arises an effective force
\be
\diff \bT_a = -\diff \bt_a \equiv \mathfrak{f}_a\,.
\ee
The interpretation of this force as an inertial effect was already discussed in Section \ref{gforces}.
In the case of CGR it turns out that $\mathfrak{f}_a$ has precisely the correct from to ensure the metric-covariant conservation of matter. This could be expected
from that the CGR can be derived from an action principle, which guarantees the generalised Bianchi identity \cite{Koivisto:2005yk}.

\subsection{Energy-momentum current}

\subsubsection{Electromagnetism}

Energy-momentum currents are described by covector-valued 3-forms. In the case of electromagnetism, we write
\be \label{emcurrent}
\prescript{\text{em}}{}\bT_a = \frac{1}{2}\lp \bF\wedge \ie_a\cdot \bH - \bH \wedge \ie_a\wedge \bF + \bA \wedge \ie_a \cdot \bP - \bP\wedge\ie_a\cdot \bA\rp\,.
\ee
If the theory can be obtained from a twisted Lagrangian 4-form
\be
\bLambda = -\frac{1}{2}\bF\wedge\bH - \frac{1}{2}\bA\wedge\bP\,,
\ee
we can alternatively write
\be
\prescript{\text{em}}{}\bT_a = \ie_a\cdot \bLambda - \bF\wedge\ie_a\cdot\bH  - \bA\wedge\ie_a\cdot\bP\,.
\ee
One may verify that, defining $\mathcal{L}_a\bX = [\ie_a,\bX]$ for any form $\bX$ as its Lie derivative along the basis vector $\ie_a$,
\be
\diff \prescript{\text{em}}{}\bT_a = \boldsymbol{f}_a - \frac{1}{2}\lp \bF\wedge\mathcal{L}_a\bH - \bH\wedge\mathcal{L}_a\bF
+ \bA\wedge\mathcal{L}_a\bP - \bP\wedge\mathcal{L}_a\bA\rp\,.
\ee
Here $\boldsymbol{f}_a$ is the Lorentz force and the remaining additional force is determined by the constitutive law.

\subsubsection{Gravity}

In case of gravity, the current analogous to (\ref{emcurrent}) contains only the two last terms. That is,
\be
\bt_a = \frac{1}{2}\lp \bA^{bc} \wedge \ie_a \cdot \bP_{bc} - \bP_{bc}\wedge\ie_a\cdot \bA^{bc}\rp\,.
\ee
Again, if the theory can be obtained from a twisted Lagrangian 4-form
\be
\bLambda =  - \frac{1}{2}\bA^{ab}\wedge\bP_{ab}\,,
\ee
there is an alternative formula,
\be
\bt_a = \ie_a\cdot \bLambda  - \bA^{bc}\wedge\ie_a\cdot\bP_{bc}\,.
\ee
The conservation of the gravitational energy-momentum tensor can be derived to be
\be
\diff \bt_a = - \frac{1}{2}\lp  \bA^{bc}\wedge\mathcal{L}_a\bP_{bc} - \bP_{bc}\wedge\mathcal{L}_a\bA^{bc}\rp\,,
\ee
which is minus the effective force $\mathfrak{f}_a$ affecting matter.

\subsection{Constitutive relation}

\subsubsection{Electromagnetism}

By using the expansions
\be
\bH = \frac{1}{2}\tilde{H}^{ab}\epsilon_{ab}\,, \quad \bF = \frac{1}{2}F_{ab}\e^{ab}\,, \quad \bP = \tilde{P}^a\epsilon_a\,, \quad \bA = A_a\e^a\,,
\ee
one finds that essentially the same constitutive relations as in subsection \ref{electroconst} are to be specified. For example, a generalisation of the Proca theory is given by
\be
\tilde{H}^{ab} = {\chi}^{abcd}F_{cd}\,, \quad \tilde{P}^a = {\xi}^{ab}A_b\,.
\ee
The constitutive tensor densities in the two languages are related by the components of the coframe field
\be
 {\chi}^{abcd} = \e^a{}_\mu\e^b{}_\nu e^c{}_\rho e^d{}_\sigma {\chi}^{\mu\nu\rho\sigma}\,, \quad {\xi}^{ab} = \e^a{}_\mu\e^b{}_\nu {\xi}^{\mu\nu}\,.
\ee
Thus, the analysis of the constitutive relations in the tensor language is directly applicable to theory formulated in the exterior algebra.

\subsubsection{Gravity}

Now let us recall the expansions
\be
\bH_{a}{}_ = \frac{1}{2}\tilde{H}_{a}{}^{bc}\epsilon_{bc}\,, \quad {\bP}_{ab} = \tilde{P}^c{}_{ab}\epsilon_c\,, \quad \bA^{ab}{} = A^{ab}{}_c\e^c\,.
\ee
The constitutive relations we focused upon previously are thus expressed in latin indices as
\be
\tilde{H}_c{}^{ab} = \chi^{ab}{}_{cdf}{}^e Q_e{}^{df}\,, \quad \tilde{P}^a{}_{bc} = \xi^a{}_{bc}{}^d{}_{ef}Q_d{}^{ef}\,,
\ee
where the index conversion can be made with the help of the components of the coframe and the tetrads, i.e. the components of the frame, in the very obvious way
\be
\chi^{ab}{}_{cdf}{}^e = \e^a{}_\mu\e^b{}_\nu\ie_f{}^\sigma\e^e{}_\beta\chi^{\mu\nu}{}_{\alpha\rho\sigma}{}^\beta \ie_c{}^\alpha\ie_d{}^\rho\,,  \quad
 \xi^a{}_{bc}{}^d{}_{ef} = \e^a{}_\alpha\e^d{}_\beta \xi^\alpha{}_{\mu\nu}{}^\beta{}_{\rho\sigma} \ie_b{}^\mu\ie_c{}^\nu\ie_e{}^\rho\ie_f{}^\sigma\,.
\ee
By lifting the analysis into the frame bundle, we have introduced an additional object, the frame field, only to avoid introducing a coordinate chart explicitly. One may then consider trading the extra structure
for another, the symbol $\eta^{ab}$. Then one of the fields, $\e^a$ and $\varphi^{ab}$ (where $\bQ^{ab}=-\diff \varphi^{ab}$) becomes redundant, since it is possible, and indeed
conventional, to make the identification $\varphi^{ab}\cdot\ie_a\cdot\ie_b = \eta^{ab}\ie_a\otimes\ie_b$.

\begin{center}
\begin{table}[h]
\begin{tabular}{| c| c c c|c c c|}
\hline
 Objects and laws & Basis & Electromagnetism & Gravity & W & Electromagnetism & Gravity  \\
\hline
 Source current & 3- & $\bJ = \bT + \bt$  & $\bJ_{a}=\bT_{a}+\bt_{a}$ & 1 &$J^\mu{}=T^\mu{}+t^\mu{}_\nu$ & $J^\mu{}_\nu=T^\mu{}_\nu+t^\mu{}_\nu$ \\
 Conservation law & 4- & $\diff \bJ=0$ & $\diff \bJ_{a}=0$   & 1 & $\nabla_\mu J^\mu=0$& $\nabla_\mu J^\mu{}_\nu=0$ \\
 Kinetic excitation & 2- & $\bH$ & $\bH_{a}$& 1 & $H^{\mu\nu}$& $H^{\mu\nu}{}_\alpha$\\
  Mass excitation & 3- & $\bP$ & $\bP_{ab}$ & 1 & $P^\alpha$ & $P^\alpha{}_{\mu\nu}$ \\
 Inhomog. field eqn. & 3- & $\diff\bH=\bJ$ & $\diff\bH_{a}=\bJ_{a}$& 1 & $\nabla_\mu H^{\mu\nu}=J^\nu$ & $\nabla_\mu H^{\mu\nu}{}_\alpha=J^\mu{}_\alpha$ \\
 Kinetic potential & 1+ & $\bA$ & $\bA^{ab}$& 0 & $A_\mu$ & $A^{\alpha\beta}{}_\nu$ \\
 Mass potential & 0+ & $B$ & $B^{ab}$& 0 & $B$ & $B^{\alpha\beta}$ \\
 Field strength & 2+ & $\bF=\diff\bA$ & $\bF^{ab}=\diff\bA^{ab}$ & $0$ & $F_{\mu\nu}=2\nabla_{[\mu}A_{\nu]}$ & $F^{\alpha\beta}{}_{\mu\nu}=2\nabla_{[\mu}A^{\alpha\beta}{}_{\nu]}$ \\
 Homog. field eqn & 3+, 2+ & $\diff \bF=0$ & $\bF^{ab}=0$ & $0$ & $\nabla_{[\alpha}F_{\mu\nu]}=0$ & $F^{\alpha\beta}{}_{\mu\nu}=0$\\
 Lorentz force & 4- & $f_a=\ie_a\cdot \bF\wedge\bJ$ & $0$ &  $1$ & $f_\mu = F_{\mu\nu}J^\nu$ & $0$ \\
 Effective force & 4- & $\boldsymbol{\mathfrak{f}}=-\diff\bt$ & $\boldsymbol{\mathfrak{f}}_{a}=-\diff\bt_{a}$& $1$& $\mathfrak{f}=-\nabla_\mu t^\mu$ & $\mathfrak{f}_\nu=-\nabla_\mu t^\mu{}_\nu$\\
 Energy-momentum & 3- & $\prescript{\text{em}}{}{\bT}_a$ & $\bt_{a}$ & $1$ & $\prescript{\text{em}}{}{T}^\mu{}_\nu$ & $t^\mu{}_\nu$ \\
 Kinetic Lagrangian & 4- & $\prescript{\text{kin}}{}\bLambda = -\frac{1}{2}\bF\wedge\bH$ & $0$ & 1 & $\prescript{\text{kin}}{}L = \frac{1}{4}H^{\mu\nu}F_{\mu\nu}$ & $0$ \\
 Mass Lagrangian & 4- & $\prescript{\text{pot}}{}\bLambda = - \frac{1}{2}\bA\wedge\bP$ & $\prescript{\text{pot}}{}\bLambda =-\frac{1}{2}\bA^{ab}\wedge\bP_{ab}$ & 1 & $\prescript{\text{pot}}{}L=-\frac{1}{2}P^\mu A_\mu$& $\prescript{\text{pot}}{}L=-\frac{1}{2}P^\alpha{}_{\mu\nu} A^{\mu\nu}{}_\alpha$ \\
 \hline
 Kinetic constitutive rel. & 0- & $\chi^{abcd}$ & $\chi^{ab}{}_{cdf}{}^e$ & 1 & $\chi^{\alpha\beta\mu\nu}$ & $\chi^{\mu\nu}{}_{\alpha\beta\gamma}{}^\delta$ \\
 Mass constitutive rel. & 0- & $\xi^{ab}$ & $ \xi^a{}_{cd}{}^b{}_{ef}$ & 1 & $\xi^{\alpha\beta}$ & $ \xi^\alpha{}_{\mu\nu}{}^\beta{}_{\rho\sigma}$ \\
 \hline
\end{tabular}
\caption{Summary of the objects and laws in the two formalisms.   In the column ``Basis'' the number $p$ denotes a $p$-form, and $-$ is for twisted, $+$ for untwisted. The entry in column ``W'' is $1$ for tensor densities (or, tensor
density equations) and $0$ for tensors (the rank is manifest). We see that the quantities corresponding to twisted forms are tensor densities. We also see that apart from the form of the homogeneous field equation, the analogy with
massive electromagnetism is complete, naturally modulo the extra indices in gravity theory. However, as mentioned in the text, the homogeneous field equation $\bF^{ab}=0$ could be regarded as a macroscopic approximation to the
gravity theory where only $\diff \bF^{ab}=0$ was required at energies $\sim M$ (or, distances $\sim 1/M$). \label{table1}}
\end{table}
\end{center}

\section{Restoring symmetries}
\label{symmetries}

In this Section we discuss further the analogy of purified gravity and massive electromagnetism. In particular, the analogy suggests a natural extrapolation of CGR which
predicts an ``impurity'' of the spacetime structure at the Plank scale.

\subsection{From Proca to Stueckelberg}
\label{ptos}

\subsubsection{Electromagnetism}

This far our discussion has been based on the Proca formulation. The Lagrangian for the massive vector field is then simply
\be \label{proca}
L_{\text{Proca}} = \frac{1}{4}F_{\mu\nu}F^{\mu\nu} - \frac{1}{2}m^2 A_\mu A^\mu\,.
\ee
The mass term obviously breaks the gauge symmetry $A_\mu \rightarrow A_\mu - \nabla_\mu\varphi$. For many purposes, it is much better to consider the Stueckelberg version of the
theory. Let us introduce a new field $B$ and
write the Lagrangian for the two fields as
\be \label{stuck}
 L_{\text{Stueck}} = \frac{1}{4}F_{\mu\nu}F^{\mu\nu} - \frac{1}{2}\lp\nabla_\mu B+m A_\mu\rp\lp\nabla^\mu B + m A^\mu\rp\,.
\ee
The trick is that now we have restored the gauge symmetry of the massless case, when taking into account also the transformation of the field $B$, since the action is invariant under
\be
A_\mu \rightarrow A_\mu - \nabla_\mu\varphi\,, \quad B \rightarrow B + m\varphi\,, 
\ee
and by setting $B=0$ we recover the Proca action. Thus, the formulations (\ref{proca}) and (\ref{stuck}) are equivalent as far as the vector boson is concerned, but the introduction of a further degree of freedom, $B$, allows to extend the symmetries of the system yielding important consequences, for instance, for the renormalizability of the theory \cite{Ruegg:2003ps}.

This suggests to improve the application of the premetric program, as realised in the two previous sections, in the case of nonzero potential excitations. It should be then understood
that just as the existence of the kinetic excitation $\bH$ implies the existence of a potential $\bA$, the existence of a potential excitation $\bP$ implies the existence of a
field $B$. The principle is that the symmetry that emerges for the kinetic excitation should not be destroyed by the presence of the potential excitation. Thus, a non-vanishing
constitutive relation $\xi$ may be considered to entail the presence of an additional Stueckelberg field. As is the case in the above demonstrated example, the resulting theory should be physically completely
equivalent (despite the formal introduction of the additional degrees of freedom). However, it is, in our understanding, conceptually preferable from the standpoint of the premetric program, to consider that the redundancy in the one-form
$\bA$ deduced from the property of the two-form $\bH$, is not undermined by the presence of the three-form $\bP$. Rather, from the latter, we can deduce the further property of the theory: the existence of the 0-form $B$.

In the case of purified gravity, we shall sometimes refer to the corresponding Stueckelberg 0-form $B^{ab}$ as the ``premetric field''.

\subsubsection{Gravity}
\label{clarity}

Two concerns may have arisen in the formulation of the gravitational theory as presented in the above three Sections. Firstly, was it legitimate to start with the connection $\nabla_\mu$ instead of $\partial_\mu$? Secondly, was it legitimate to promote the gauge transformation $\varphi^{\mu\nu}$ (respectively, $\varphi^{ab}$) to a dynamical variable? Now we shall clarify these points which had been
left somewhat vague. Both the issues are addressed by applying the same symmetry-based reasoning that lead us from Proca electromagnetism to Stuckelberg
electromagnetism to gravity. Thus, neither the substitution $\partial_\mu \rightarrow \nabla_\mu$, nor the substitution $A^{\mu\nu}{}_\alpha \rightarrow -Q_\alpha{}^{\mu\nu} =\nabla_\alpha g^{\mu\nu}$
are put by hand in the theory, but they are the inevitable consequences of the symmetry axiom we proposed to supplement the premetric program with in order to more robustly deal with a
symmetry-breaking self-interactions when such arise in physics.

To show this in detail, let us take some steps back and undo those two perhaps dubious substitutions.
In the Proca-type formulation we have now (twice) walked through, the gravitational Lagrangian is written as
\be \label{procag}
L_{P} = \lambda_{\mu\nu}{}^{\alpha\beta} F^{\mu\nu}{}_{\alpha\beta} +
\frac{1}{2}M^2A^{\mu\nu}{}_\alpha \xi^\alpha{}_{\mu\nu}{}^\beta{}_{\rho\sigma} A^{\rho\sigma}{}_{\beta}\,,
\ee
where we implemented the constraint of vanishing force with a Lagrange multiplier $\lambda_{\mu\nu}{}^{\alpha\beta}$ (though this is inessential). The important thing is that we do not assume
anything else about the field $A^{\rho\sigma}{}_{\beta}$, except that it does not describe a physical force. The symmetry of the field strength $F^{\mu\nu}{}_{\alpha\beta}$ for the gauge potential $A^{\mu\nu}{}_\alpha$ is
$A^{\mu\nu}{}_\alpha \rightarrow A^{\mu\nu}{}_\alpha - \partial_\alpha\varphi^{\mu\nu}$ for an arbitrary $\varphi^{\alpha\beta}$. Obviously, the self-interaction term now breaks such symmetry.

Exactly as in the case of electromagnetic self-interaction, we shall restore the symmetry by
introducing the compensating field $B^{\mu\nu}$, which must transform as $B^{\mu\nu} \rightarrow B^{\mu\nu} + M\varphi^{\mu\nu}$. The gravitational Stueckelberg action then reads
\be
L_{S} = \lambda_{\mu\nu}{}^{\alpha\beta} F^{\mu\nu}{}_{\alpha\beta} +
\frac{1}{2}\lp M A^{\mu\nu}{}_\alpha + \partial_\alpha B^{\mu\nu}\rp \xi^\alpha{}_{\mu\nu}{}^\beta{}_{\rho\sigma}\lp M A^{\rho\sigma}{}_\beta +\partial_\beta B^{\rho\sigma}\rp\,.
\ee
The variation with respect to the field $A^{\mu\nu}{}_\alpha$ just gives an equation of motion for the irrelevant Lagrange multiplier:
\be
2\partial_\beta \lambda_{\mu\nu}{}^{\alpha\beta} = \xi^\alpha{}_{\mu\nu}{}^\beta{}_{\rho\sigma}\lp M A^{\rho\sigma}{}_\beta +\partial_\beta B^{\rho\sigma}\rp\,.
\ee
The variation with respect to the Lagrange multiplier in turn, gives the equation of motion for $A^{\mu\nu}{}_\alpha$:
\be
F^{\mu\nu}{}_{\alpha\beta} = 0 \quad \Rightarrow \quad A^{\mu\nu}{}_\alpha = 2M\hat{\Gamma}^{\mu}{}_\alpha{}^{\nu}\,, \quad \text{where}\,\, \hat{\Gamma}^{\mu}{}_\alpha{}^{\nu}\,\,\text{is flat}\,.
\ee
We have introduced a flat affine connection with a convenient normalisation.
Using this information in the action, it becomes
\be
L_{S} =
\frac{1}{2}M^2\lp  \partial_\alpha B^{\mu\nu} + 2\hat{\Gamma}^{(\mu}{}_\alpha{}^{\nu)} \rp \xi^\alpha{}_{\mu\nu}{}^\beta{}_{\rho\sigma}\lp \partial_\beta B^{\rho\sigma} +  2\hat{\Gamma}^{(\rho}{}_\alpha{}^{\sigma)} \rp\,.
\ee
We have emphasised the symmetrisation of the connection $\hat{\Gamma}^{\mu}{}_\alpha{}^{\nu}$ that is imposed by the symmetries of the constitutive relation, $\xi^\alpha{}_{\mu\nu}{}^\beta{}_{\rho\sigma}
= \xi^\alpha{}_{(\mu\nu)}{}^\beta{}_{(\rho\sigma)}$. We should now note that though torsion is given by the antisymmetry of the last two indices of an affine connection, the contortion (i.e. the total contribution of the torsion to the affine connection) is
antisymmetric in its first and the last indices. Thus we may write
\be
\partial_\alpha B^{\mu\nu} + 2\hat{\Gamma}^{(\mu}{}_\alpha{}^{\nu)} = \partial_\alpha B^{\mu\nu} + 2{\Gamma}^{(\mu}{}_\alpha{}^{\nu)} \equiv \nabla_\alpha B^{\mu\nu}\,, \quad \text{where}\,\, {\Gamma}^{\mu}{}_\alpha{}^{\nu}\,\,\text{is flat and torsion-free}\,.
\ee
Finally we may rename the variable $B^{\mu\nu}$ as  $g^{\mu\nu}$, and conclude that the theory is
\be
L_S = \frac{1}{2}M^2\nabla_\alpha g^{\mu\nu}{}_\alpha \xi^\alpha{}_{\mu\nu}{}^\beta{}_{\rho\sigma} \nabla_\beta g^{\rho\sigma} = \frac{1}{2}M^2 Q_\alpha{}^{\mu\nu}P^\alpha{}_{\mu\nu}\,.
\ee
Thus, neither of the two ingredients of CGR (and its generalisations), the metric and the symmetric teleparallel covariant derivative\footnote{In teleparallel gravity, the relevance of the pure-gauge Lorentz connection has been emphasised often \cite{Aldrovandi:2013wha,Golovnev:2017dox,Krssak:2018ywd,Jarv:2019ctf}. It would seem more difficult to justify this structure from the premetric approach.}, are in the least way ad hoc. They emerge as dynamical variables from the first principles of the axiomatic approach to gravity theory.

\subsection{From Stueckelberg to Kibble ?}

\subsubsection{Electromagnetism}

In the formulation that was originally due to Kibble, the Stueckelberg theory was embedded in a free Abelian Higgs model. A complex scalar field $\Phi$, charged under the U(1)
symmetry, was introduced such that its U(1)-covariant derivative is
\be
\Diff_\mu \Phi = \lp\partial_\mu -ieA_\mu\rp\Phi\,.
\ee
The quite elegant action then reads
\be \label{kibble}
L_{\text{Kibble}} = \frac{1}{4}F_{\mu\nu}F^{\mu\nu} - \frac{1}{2}| \Diff \Phi |^2 + V(|\Phi |^2)\,.
\ee
A suitable potential can lead to a spontaneous symmetry breaking which sets the modulus of the complex scalar field to $|\Phi_0 | = m/e$.
Thus, at the minimum,
\be
\Phi_0 = \frac{m}{e}\exp{\lp\frac{ieB(x)}{m}\rp}\,,
\ee
and we can identify the phase of the complex scalar with the Stueckelberg field $B$. Up to a possibly nonzero $V(|\Phi_0 |^2)$, we obviously recover (\ref{stuck}) at the minimum.

\subsubsection{Gravity}

Since as far as we know, all massive elementary particles and gauge fields have acquired their mass via the Higgs mechanism, it is reasonable to assume that this is the case also for the mass $M$ of the
gravitational gauge field $A^{\alpha\beta}{}_\mu$. In fact, there is plenty of evidence for scale invariance in physics. Starting from the basic argument of Weyl \cite{weyl1919raum} which many still find compelling enough by itself
\cite{Scholz:2017pfo}, in modern particle physics technical arguments have kept suggesting the non-existence of an absolute scale at the level of fundamental physics. In a scale-invariant theory, dimensional regularisation introduces only logarithmic runnings of the coupling constants,
and in such a case the large hierarchy of the interactions might be better explained \cite{Meissner:2006zh}. If the scale symmetry was restored in the absence of the cosmological constant, its
exceedingly small value would be technically natural \cite{Lucat:2018slu}. The Higgs mass parameter sits at the edge of the stability bound \cite{Alekhin:2012py} and its quartic coupling seems to
run to zero near the Planck scale, which suggests that scale symmetry is indeed restored there if not at lower scales, possibly solving the stability issue of the Standard Model \cite{Oda:2015gna}.

For the reasons stated above, we believe the Planck scale should be the result of a spontaneous symmetry breaking. The most straightforward analogy of Kibble's Abelian Higgs model does not however satisfactorily incorporate such
a mechanism within our present formalism. We could write the Kibble-type action for gravity as
\be
L_K = \frac{1}{2}\lp\Diff_\alpha\Phi^{\mu\nu}\rp^\ast\xi^\alpha{}_{\mu\nu}{}^\beta{}_{\rho\sigma}\lp\Diff_\beta\Phi^{\rho\sigma}\rp + V_{\alpha\beta}\Phi^{\alpha\beta}\,.
\ee
To remain in the premetric framework, we have taken the potential term to be determined by the (generally nonlinear) constitutive relation $V_{\alpha\beta}$ for our new complex field
$\Phi^{\alpha\beta}$.
When the radial component of the field has settled to an isotropic constant value such that
\be
\Phi_0^{\alpha\beta} = M\exp{\lp iB^{\alpha\beta}\rp}\,,
\ee
we recover the Stueckelberg-type action. However, either we would have to invoke an inhomogeneous covariant derivative for the field such that
\be
\Diff_\mu\Phi^{\alpha\beta} = \partial_\mu \Phi^{\alpha\beta} + 2iM A^{\alpha\beta}{}_\mu{}\,,
\ee
or introduce a metric such that we would be allowed to write
\be
\hat{\nabla}_\mu\Phi^{\alpha\beta} = \partial_\mu \Phi^{\alpha\beta} + i\hat{\Gamma}^{\alpha}{}_{\mu\rho}\Phi^{\rho\beta} + i\hat{\Gamma}^{\beta}{}_{\mu\rho}\Phi^{\alpha\rho}\,.
\ee
On the the other hand, in any case it is difficult to see how now to specify the constitutive relation in practice without invoking a metric.

As the purpose of this paper was only the axiomatic formulation of purified gravity, we leave the intriguing problem of the actual symmetry breaking mechanism and its dynamics to a further study.
Recently, a promising way has been pointed out, first by realising an observer space \cite{Gielen:2012fz} in Cartan geometry \cite{Westman:2014yca} by using merely a Lorentz connection and a Higgs-like scalar (in particular: no metric or a frame field) in a polynomial quadratic action (thus, not only a local and linear but even a polynomial constitutive relation) to give rise to a spacetime in the spontaneously broken phase \cite{Zlosnik:2018qvg}, and then by embedding this scenario to the General Linear bundle \cite{Koivisto:2019ejt}, thus bringing it a step closer to our present premetric construction of purified gravity\footnote{As a side-product, this scheme may eliminate the need for particle dark matter in cosmology and astrophysics \cite{Zlosnik:2018qvg}, and suggests the incorporation of the gauge fields of particle physics within the $A^{\alpha\beta}{}_\mu$ \cite{Koivisto:2019ejt}.}.

\subsection{Impurities near the Planck scale ?}

There is a complete analogy between purified gravity and massive electromagnetism, except that the gravitational force is imposed to vanish. This raises the question why should we not allow a kinetic term for
gravity as well, and then to recover the predictions of GR, restrict to solutions with vanishing field strength. Actually, such solutions are naturally the relevant ones at the classical level. Since the connection
$A^{\mu\nu}{}_\alpha$ is massive, it interacts only at finite distances. The range of the force is of the order of the Planck length, about $1.6\cdot 10^{-35}$ meters. At the macroscopic scales, the gauge field does not
propagate. Therefore, for practical purposes $F^{\mu\nu}{}_{\alpha\beta} \approx 0$ and we obtain the same predictions by considering the theory with the canonical kinetic term, instead of imposing the vanishing of the
field strength with a Lagrange multiplier. It is interesting to consider that purified gravity becomes impure as one probes microscopic distances approaching the Planck scale. We would thus predict that at those scales, gravity
is no longer pure inertia, and the equivalence principle may in some sense be broken.

There are some similarities with this scenario and the very interesting approach towards quantum gravity by Percacci {\it et al} \cite{Percacci:1990wy,Percacci:2009ij,Pagani:2015ema}. In particular, they also consider a Higgs-like
mechanism that would make the gravitational connection massive. However, their set-up is also fundamentally different since they consider the usual (equivalent of) Einstein-Hilbert term as the kinetic term, while the mass terms are
additional terms that give masses only to the distorted part of the connection (torsion and non-metricity). In the version of purified gravity advocated in this paper, we on the contrary see the (equivalent of the) Einstein-Hilbert
term as the mass term, and speculate further that at extremely high energies the possible kinetic term, which as usual would be a square of the field strength of the connection, would become dynamically relevant.
In general, there are many interesting studies that implement the idea of a gravitational Higgs mechanism \cite{Percacci:1990wy,Percacci:2009ij,Pagani:2015ema,Isham:1971dv,Tresguerres:2000qn,Leclerc:2005qc,Tiemblo:2005js,Ali:2007hu,Westman:2014yca,Zlosnik:2018qvg,Koivisto:2019ejt}, but to our knowledge, the necessity for such a mechanism had not been previously deduced by an axiomatic method in the framework of the premetric program.

To be concrete, we propose the extrapolation of purified gravity Lagrangian $L_{\text{gr}}$ motivated by the massive electromagnetic Lagrangian $L_{\text{em}}$ like so,
(where for clarity we divide the corresponding mass scales in the $P$'s out from the $p$'s)
\be \label{two}
L_{\text{em}} = \frac{1}{4}H^{\mu\nu}F_{\mu\nu} - \frac{1}{2}m^2 p^\mu A_\mu\,, \quad
L_{\text{gr}} = \frac{1}{4}H_{\alpha\beta}{}^{\mu\nu} F^{\alpha\beta}{}_{\mu\nu} -
\frac{1}{2}M^2p_{\alpha\beta}{}^\mu A^{\alpha\beta}{}_{\mu}\,.
\ee
The hypothetical photon mass has not been detected despite conducting experiments with exquisite accuracy, and thus if an $m\neq 0$ exists, we know that this $m$ must be very small \cite{Tu:2005ge,Goldhaber:2008xy}.
On the other hand, the hypothetical ``hypercurvature'' has not been experimentally probed because, as we know, $M$ is very large, in fact the largest fundamental mass scale known in Nature.
In this sense, the phenomenological status of the two theories in (\ref{two}) are the opposite: in electromagnetism, it is only the gauge-invariant piece that appears
in the standard theory and the physicality of the longitudinal polarisation $\varphi$ has not been established, whereas for gravity it is only the invariance-breaking piece $\varphi^{\alpha\beta}$
that propagates the well-established graviton, while the gauge-invariant kinetic term of the ``hypergravitational'' gauge field $A^{\alpha\beta}{}_\mu$ is only suggested by
the theoretical argument we have put forward.

\section{Properties of general theories}
\label{properties}

In this Section we investigate the implications of more general constituve relations. First we consider the dispersion relation in a fully general (local and linear) case, and then
focus on the properties of the 13-parameter theories that can be defined in the presence of a metric.

\subsection{Wave propagation}

In gravity the inhomogeneous field equation~\eqref{gravity} has two contributions to the current \(J^{\mu}{}_{\alpha}\), which are the matter energy-momentum \(T^{\mu}{}_{\alpha}\) and gravitational energy-momentum \(t^{\mu}{}_{\alpha}\). In the geometric optics approximation we assume that the gravitational field is sufficiently weak so that we can neglect its energy-momentum contribution, and so we will set \(t^{\mu}{}_{\alpha} = 0\). Further, we study the propagation of waves in vacuum only, and hence set \(T^{\mu}{}_{\alpha} = 0\) as well. We end up with the source-free field equation \(\nabla_{\mu}H^{\mu\nu}{}_{\alpha} = 0\). We then make use of the constitutive relation~\eqref{gravconst}, together with \(Q_{\beta}{}^{\rho\sigma} = -\nabla_{\beta}\varphi^{\rho\sigma}\) and \(\varphi^{[\rho\sigma]} = 0\). Working in the Fourier domain, where \(\nabla_{\mu} \rightarrow q_{\mu}\) becomes the wave covector, we finally obtain the dispersion relation
\begin{equation}
M^{\nu}{}_{\alpha\rho\sigma}\varphi^{\rho\sigma} = 0\,, \quad
M^{\nu}{}_{\alpha\rho\sigma} = q_{\mu}q_{\beta}\chi^{\mu\nu}{}_{\alpha\rho\sigma}{}^{\beta}\,.
\end{equation}
We call \(M^{\nu}{}_{\alpha\rho\sigma}\) the characteristic tensor density. Note that not all of the irreducible components of \(\chi^{\mu\nu}{}_{\alpha\rho\sigma}{}^{\beta}\) contribute to the characteristic tensor density and hence the dispersion relation. Following their definition~\eqref{eq:chidecomp}, the components \(\prescript{[3]}{}\chi^{\mu\nu}{}_{\alpha\rho\sigma}{}^{\beta}\) and \(\prescript{[4]}{}\chi^{\mu\nu}{}_{\alpha\rho\sigma}{}^{\beta}\) are antisymmetric in their first and last indices, so that their contribution vanishes.

We further remark that we have found 16 homogeneous, linear equations for the 10 components of \(\varphi^{\rho\sigma}\). This means that there must be a redundancy which eliminates six of these equations. Four equations are readily eliminated realizing that \(q_{\nu}M^{\nu}{}_{\alpha\rho\sigma} = 0\), due to the antisymmetry of \(\chi^{\mu\nu}{}_{\alpha\rho\sigma}{}^{\beta}\) in its first two indices. The remaining redundancies are more difficult to find and depend on the particular form of the constitutive density. We will reveal them in the most general metric case below.

\subsection{Wave propagation in the metric case}

For the metric constitutive density~(\ref{eq:chinewer},\ref{eq:chiodd}) we find the characteristic tensor density
\begin{equation}
M_{\nu\alpha\rho\sigma} = \frac{\sqrt{-g}}{2}\left[b_1\left(q_{\nu}g_{\alpha(\rho}q_{\sigma)} - q^2g_{\nu(\rho}g_{\sigma)\alpha}\right) + b_2\left(q_{\nu}q_{\alpha} - q^2g_{\nu\alpha}\right)g_{\rho\sigma} + b_3\left(q_{\alpha}g_{\nu(\rho}q_{\sigma)} - g_{\nu\alpha}q_{\rho}q_{\sigma}\right)\right] + \frac{2b_4 - b_6}{2}\epsilon_{\nu\alpha\beta(\rho}q_{\sigma)}q^{\beta}\,,
\end{equation}
where we have lowered the first index for convenience, and introduced the abbreviation \(q^2 = q_{\mu}q^{\mu}\). We see that the terms corresponding to the parameters \(b_5\) and \(b_7\) do not contribute. The linearized field equations thus read
\begin{multline}\label{eq:wavefull}
0 = E_{\nu\alpha} = 2M_{\nu\alpha\rho\sigma}\varphi^{\rho\sigma}\\
=\sqrt{-g}\left[b_1(\varphi_{\alpha\beta}q^{\beta}q_{\nu} - q^2\varphi_{\nu\alpha}) + b_2\varphi^{\beta}{}_{\beta}(q_{\nu}q_{\alpha} - q^2g_{\nu\alpha}) + b_3(\varphi_{\nu\beta}q^{\beta}q_{\alpha} - g_{\nu\alpha}\varphi^{\rho\sigma}q_{\rho}q_{\sigma})\right] + (2b_4 - b_6)\epsilon_{\nu\alpha\beta\rho}q^{\beta}q_{\sigma}\varphi^{\rho\sigma}\,.
\end{multline}
As mentioned before, these equations are not independent, and the four equations \(q^{\nu}E_{\nu\alpha}\) are satisfied identically. To find further redundancies, it is helpful to further decompose these equations into four irreducible parts. For this purpose, we first contract with \(q^{\alpha}\), which yields the longitudinal part
\begin{equation}\label{eq:wavelong}
E_{\nu\alpha}q^{\alpha} = 2\sqrt{-g}(b_1 - b_3)q^{\alpha}q^{\beta}\varphi_{\beta[\alpha}q_{\nu]}\,.
\end{equation}
Remarkably, the antisymmetric, transverse part
\begin{equation}\label{eq:waveanti}
E^{\beta\gamma}\epsilon_{\beta\gamma\nu\alpha}q^{\alpha} = 4(2b_4 - b_6)q^{\alpha}q^{\beta}\varphi_{\beta[\alpha}q_{\nu]}
\end{equation}
is of the same form, but is the only part which originates from the parity-violating terms. The trace of the field equations reads
\begin{equation}\label{eq:wavetrace}
E^{\alpha}{}_{\alpha} = \sqrt{-g}[(b_1 - 3b_3)\varphi^{\alpha}{}_{\alpha}q^2 - (b_1 + 3b_2)\varphi^{\alpha\beta}q_{\alpha}q_{\beta}]\,.
\end{equation}
We are then left with the transverse and trace-free part of the field equations, which reads
\begin{multline}\label{eq:wavesym}
3q^2E_{(\nu\alpha)} - 3q^{\beta}q_{(\nu}E_{\alpha)\beta} + E^{\beta}{}_{\beta}(q_{\nu}q_{\alpha} - q^2g_{\nu\alpha}) =\\
\sqrt{-g}b_1\left\{6q^2q^{\beta}q_{(\nu}\varphi_{\alpha)\beta} - q_{\nu}q_{\alpha}(q^2\varphi^{\beta}{}_{\beta} + 2q_{\beta}q_{\gamma}\varphi^{\beta\gamma}) - 3(q^2)^2\varphi_{\nu\alpha} + [(q^2)^2\varphi^{\beta}{}_{\beta} - q^2q_{\beta}q_{\gamma}\varphi^{\beta\gamma}]q_{\nu\alpha}\right\}
\end{multline}
and, again remarkably, depends on \(b_1\) only. Note that each of these equations must be satisfied individually.

We now find the aforementioned redundancy of the field equations, which is apparent from the fact that the longitudinal part~\eqref{eq:wavelong} and the antisymmetric transverse part~\eqref{eq:waveanti} are identical, up to a constant factor, which means that only one of them counts to the number of independent equations. These are four equations since their is one free index; however, they are not independent, since their contraction with \(q^{\nu}\) vanishes identically. Hence, we keep three equations, and have eliminated three further redundant equations, in addition to the four equations already found for the general constitutive relation. In total we have thus eliminated seven of the original 16 equations. The remaining nine equations are the trace equation~\eqref{eq:wavetrace}, the five independent components of the symmetric, transverse, trace-free equation~\eqref{eq:wavesym} and the three independent components mentioned before. Since \(\varphi^{\rho\sigma}\) has 10 independent components, it thus follows that there must be a gauge freedom eliminating one of them. This can most easily be seen from an Ansatz of the form
\begin{equation}
\varphi^{\rho\sigma} = Ug^{\rho\sigma} + Vq^{\rho}q^{\sigma} + W^{(\rho}q^{\sigma)} + \tilde{\varphi}^{\rho\sigma}\,,
\end{equation}
where
\begin{equation}
W^{\rho}q_{\rho} = 0\,, \quad
\tilde{\varphi}^{[\rho\sigma]} = 0\,, \quad
\tilde{\varphi}^{\rho}{}_{\rho} = 0\,, \quad
\tilde{\varphi}^{\rho\sigma}q_{\sigma} = 0\,.
\end{equation}
Inserting this ansatz into the field equations~\eqref{eq:wavefull}, we find that they reduce to
\begin{equation}\label{eq:wavedecom}
\sqrt{-g}\left\{2\left[(b_1 + 4b_2 + b_3)U + (b_2 + b_3)q^2V\right](q_{\alpha}q_{\nu} - q^2g_{\alpha\nu}) - \frac{b_1 - b_3}{2}q^2q_{\alpha}W_{\nu} - b_1q^2\tilde{\varphi}_{\nu\alpha}\right\} - \frac{2b_4 - b_6}{2}\epsilon_{\alpha\nu\rho\sigma}q^2q^{\rho}W^{\sigma} = 0\,,
\end{equation}
while the decomposed equations take the form
\begin{align}
\sqrt{-g}q^2[(b_1 + 4b_2 + b_3)U + (b_2 + b_3)q^2V] &= 0\,, &
\sqrt{-g}(b_1 - b_3)(q^2)^2W_{\alpha} &= 0\,, \nonumber\\
\sqrt{-g}b_1(q^2)^2\tilde{\varphi}_{\alpha\beta} &= 0\,, &
(2b_4 - b_6)(q^2)^2W_{\alpha} &= 0\,,
\end{align}
up to constant, numerical factors. We see that the scalar, vector and tensor modes decouple and that the equations~\eqref{eq:wavedecom} possess the gauge freedom
\begin{equation}
U \rightarrow U + \lambda(b_2 + b_3)q^2\,, \quad
V \rightarrow V - \lambda(b_1 + 4b_2 + b_3)\,,
\end{equation}
and hence
\begin{equation}
\varphi^{\rho\sigma} \rightarrow \varphi^{\rho\sigma} + \lambda[(b_2 + b_3)q^2g^{\rho\sigma} - (b_1 + 4b_2 + b_3)q^{\rho}q^{\sigma}]\,.
\end{equation}
This removes one of the two scalar degrees of freedom. Also for the second scalar mode we find that it is not propagating, since the corresponding terms in the field equations~\eqref{eq:wavedecom} take the form of a constraint equation. Further, we find that non-trivial solutions for the remaining modes are obtained only for \(q^2 = 0\), so that all wave solutions must propagate along the null directions of the metric \(g_{\alpha\beta}\), i.e., on its light cone.

We remark that a particular case is given by theories whose parameters satisfy \(b_1 = b_3\) and \(2b_4 = b_6\). In this case the field equations~\eqref{eq:wavefull} and hence also~\eqref{eq:wavedecom} are symmetric and the vector mode \(W^{\rho}\) does not contribute. This is in particular the case for CGR. We thus find that the only propagating mode is the transverse, traceless tensor mode, as expected.

\subsection{Perturbations}

Consider the perturbations $\delta g_{\mu\nu}$ of the flat metric $\eta_{\mu\nu}$,
\be
g_{\mu\nu} = \eta_{\mu\nu} + \delta g_{\mu\nu}\,.
\ee
Using the 1+3 decomposition familiar from cosmological perturbation theory, we decompose the perturbation $\delta g_{\mu\nu}$ into scalars $\phi$, $\psi$, $\beta$, $\sigma$, transverse vectors $B_i$, $E_i$, and transverse and traceless tensors $h_{ij}$ as follows:
\be
\delta g_{00} = -2\phi\,, \quad \delta g_{0i} = -\beta_{,i} + B_i\,, \quad \delta g_{ij} = -2\psi\delta_{ij} + \sigma_{,ij}-\frac{1}{3}\nabla^2\sigma\delta_{ij} + 2 E_{(i,j)} + 2h_{ij}\,.
\ee
We shall compute the linearised field equations in vacuum. Since $t^\mu{}_\nu$ is of quadratic order, it is sufficient to consider the equation $\nabla_\alpha H^{\alpha\mu}{}_\nu=0$.

The energy-momentum could be computed from
\ba
H^{i0}{}_0 & = &  -\lp b_1+b_2\rp \phi^{,i} + \lp 3b_2+b_3\rp\psi^{,i} + \frac{1}{2}\lp b_1-b_3\rp\dot{\beta}^{,i} +
\frac{1}{3}b_3\nabla^2\sigma^{,i} \nn \\ & + & \frac{1}{2}\lp b_1-b_3\rp \dot{B}^i + \frac{1}{2}\lp b_6-2b_7\rp\epsilon^{ijk}B_{j,k} + \frac{1}{2}b_3\nabla^2 E^i\,, \\
H^{i0}{}_j & = & -\delta^i_j\lp b_2+b_3\rp\dot{\phi} + \delta^i_j\lp b_1+3b_2\rp\dot{\psi}
+\lp 2b_5+b_6\rp\epsilon^i{}_j{}^k\phi_{,k}-\lp 2b_4+6b_5+b_6+2b_7\rp\epsilon^i{}_j{}^k\psi_{,k}   \nn \\ & + & \frac{1}{2}b_1\partial^i\partial_j\beta - \frac{1}{2}b_3\delta^i_j\nabla^2\beta -\frac{1}{2} \lp 2b_4 - b_6\rp\epsilon^i{}_j{}^k\dot{\beta}_k  \nn \\
& - & \frac{1}{2}b_1\lp \partial^i\partial_j-\frac{1}{3}\delta^i_j\nabla^2\rp\dot{\sigma} + \frac{1}{6}\lp 4b_4-b_6-2b_7\rp\epsilon^i{}_j{}^k\nabla^2\sigma_{,k} \nn \\
& + & \frac{1}{2}b_1 B_j{}^{,i} - b_4\epsilon^i{}_j{}^k\dot{B}_k - b_1\dot{E}^{(i}{}_{,j)} + b_4\epsilon^i{}_j{}^k\nabla^2 E_k - \frac{1}{2}b_6\epsilon_j{}^{kl}E_{k,l}{}^{,i}
+ b_7\epsilon^{ikl}E_{k,jl} \nn \\
& - & b_1 \dot{h}^i{}_j - b_6 \epsilon_j{}^{kl}h^i{}_{k,l} + 2b_7\epsilon^{ikl}h_{jk,l}\,.
\ea
The field equations are $\nabla_\alpha H^{\alpha\mu}{}_\nu=0$, where
\ba
\nabla_\mu H^{\mu 0}{}_0 & = &  -\lp b_1+b_2\rp\nabla^2\phi + \lp 3b_2+b_3\rp\nabla^2\psi  + \frac{1}{2}\lp b_1-b_3\rp\nabla^2\dot{\beta}
-\frac{1}{3}b_3\nabla^4\sigma\,, \label{efe1} \\
\nabla_\mu H^{\mu 0}{}_i & = & -\lp b_2+b_3\rp\dot{\phi}_{,i} + \lp b_1+3b_2\rp\dot{\psi}_{,i}   + \frac{1}{2}\lp b_1-b_3\rp\nabla^2\beta^{,i} -\frac{1}{3}b_1\nabla^2\dot{\sigma}_{,i} \nn \\
& + & \frac{1}{2}b_1 \nabla^2\lp B_i -\dot{E}_i\rp - \lp b_4 - \frac{1}{2}b_6\rp\lp \epsilon_i{}^{kl}B_{k,l}  - 2\nabla^2 E_{(j,k)}\rp \label{efe2}
\,, \\
\nabla_\mu H^{\mu i}{}_j & = & \lp b_2+b_3\rp\delta^i_j\ddot{\phi}+ b_2\lp \partial^i\partial_j -\delta^i_j\nabla^2\rp\phi - \lp b_1 + 3b_2 \rp\delta^i_j\ddot{\psi} - \lp b_1+3b_2+b_3\rp\lp\partial^i\partial_j-\delta^i_j\nabla^2\rp\psi \nn \\  & + & \lp 2b_4-b_6\rp\epsilon^i{}_j{}^k\lp\dot{\phi}_{,k}+\dot{\psi}_{,k}\rp \nn \\
& - & \frac{1}{2}\lp b_1+b_3\rp\partial^i\partial_j \dot{\beta}+ b_3\delta^i_j\nabla^2\dot{\beta} - \lp b_4-\frac{1}{2}b_6\rp \epsilon^i{}_j{}^k\lp\ddot{\beta}_{,k} + \nabla^2\beta_{,k}\rp  \nn \\
& - &  \frac{1}{6}\lp b_1-2b_3\rp\lp \partial^i\partial_j -\delta^i_j\nabla^2\rp\nabla^2\sigma
 + \frac{1}{6}b_1\lp 3\partial^i\partial_j - \delta^i_j\nabla^2\rp\ddot{\sigma} - \frac{1}{3}\lp 2b_4-b_6\rp\epsilon^i{}_j{}^k\nabla^2\dot{\sigma}_{,k} \nn \\
& - & \frac{1}{2}\lp b_1\dot{B}^{i}{}_{,j} + b_3\dot{B}_j{}^{,i}\rp -\frac{1}{2}\lp b_1-b_3\rp \nabla^2 E_j{}^{,i} + b_1\ddot{E}^i{}_{,j} + \lp b_4-\frac{1}{2}b_6\rp \epsilon^i{}_j{}^k\lp\ddot{B}_k  - \nabla^2 \dot{E}_k\rp     \nn \\     & - & b_1 \Box h^i{}_j\,. \label{efe3}
\ea
Note that these in general have also antisymmetric components, given as
\ba
\lp\nabla_\mu H^{\mu [0}{}_j\rp g^{i]j} & = & \frac{1}{2}\lp b_1-b_3\rp\lp \dot{\phi}^{,i}+\dot{\psi}^{,i}+ \ddot{\beta}^{,i} +  \nabla^2\beta^{,i}-\frac{1}{3}\nabla^2\dot{\sigma}^{,i}  + \ddot{B}^i -\nabla^2 \dot{E}^i \rp \nn \\
& - &  \lp 2b_4-b_6\rp\lp\epsilon^{ikl}B_{k,l}  - 2\nabla^2 E_{(j,k)}\rp  \,, \\
\lp\nabla_\mu H^{\mu [i}{}_j \rp g^{j]k} & = & \lp 2b_4-b_6\rp \epsilon^{ijk}\lp \dot{\phi}_{,k}+\dot{\psi}_{,k}  + \frac{1}{2}\ddot{\beta}_{,k} +  \frac{1}{2}\nabla^2\beta_{,k} -\frac{1}{3}\nabla^2\dot{\sigma}_{,k} + \ddot{B}_k  - \nabla^2 \dot{E}_k\rp \nn \\
& + &  \lp b_1-b_3\rp\lp \dot{B}^{[i,j]}-\nabla^2 E^{[i,j]}   \rp\,.
\ea
Since the scalars, vectors and tensors are decoupled at the linear order, we can focus on each sector separately.

The transverse-traceless perturbations $h_{ij}$ are the simplest. We see that as long as $b_1 \neq 0$, there are tensor perturbations that propagate on the light cone, obeying the usual wave equation $\Box h_{ij}=0$.

As expected from the analysis of the characteristic equation, the parameters $b_5$ and  $b_7$ do not enter the field equations. We see that the antisymmetric components of the field equations vanish when the longitudinal and the antisymmetric transverse part of characteristic equation are set to zero. In the following we will consider only the subset of constitutive relations which yield symmetric field equations. Thus we set $b_3=b_1$ and $b_6=2b_4$.

Then the equations of motion for the vector perturbations reduce to $\nabla^2 V^i=0$, $\dot{V}^i=0$, where $V^i=B^i-\dot{E}^i$ is the gauge-invariant combination of the two transverse 3-vectors. These equations are the same as in General Relativity. Thus, we find that vector perturbations do not propagate in vacuum.

To study the system of four coupled scalar perturbations, let us consider the Fourier modes with frequency $q^0=\omega$ and wavevector $q^i=k^i$.
One readily sees that then the two equations (\ref{efe1}) and (\ref{efe2}) become redundant.
Using one of them, the trace and the off-diagonal part of (\ref{efe3}), respectively, we obtain the three equations,
\ba
0 & = & \lp b_1+b_2\rp \phi - \lp b_1+3b_2\rp \psi + \frac{1}{3}b_1 \hat{\sigma}\,, \\
0 & = & -3\lp b_1+b_2\rp\omega^2\phi + 2b_2 k^2\phi + 3\lp b_1+3b_2\rp\omega^2\psi - 2\lp 2b_1+3b_2\rp k^2\psi -2b_1 k^2\hat{\beta} + \frac{1}{3}b_1 k^2\hat{\sigma}\,, \\
0 & = & b_2\phi-\lp 2b_1+3b_2\rp\psi - b_1\hat{\beta} + \frac{1}{6}b_1\hat{\sigma} + \frac{1}{2}b_1\frac{\omega^2}{k^2}\hat{\sigma}\,,
\ea
where we defined $\hat{\beta}=i\omega\beta$, $\hat{\sigma}=-k^2\sigma$. However, only two of the three equations above are independent. Thus we have only two equations for
four variables. This, nevertheless, is sufficient because there are now two gauge invariances, say $X$ and $Y$,
\ba
\phi & \rightarrow  & \phi - \frac{b_1+3b_2}{2\lp b_1+2b_2\rp}X\,, \quad  \psi \rightarrow \psi - \frac{b_1+b_2}{2\lp b_1+2b_2\rp}X\,, \quad \beta \rightarrow \beta + X\,, \\
\phi & \rightarrow  & \phi +\frac{\lp b_1+3b_2\rp\omega^2-\lp b_1+b_2\rp k^2}{4\lp b_1+2b_2\rp k^2} Y\,, \quad
\psi \rightarrow \psi +\frac{3\lp b_1+b_2\rp\omega^2+\lp b_1-b_2\rp k^2}{12\lp b_1+2b_2\rp k^2} Y\,, \quad
\hat{\sigma} \rightarrow \hat{\sigma} + Y\,.
\ea
We can therefore eliminate any two of the variables, and solve for the rest from the above system. If $b_1=0$, the system is underdetermined, but otherwise we find a trivial dispersion relation, i.e. no propagating scalar modes in vacuum.

In summary, the five-parameter class of theories with $b_3=b_1$ and $b_6=2b_4$ has the same field content in vacuum as CGR, and thus, to the leading order, this class of theories
is perfectly viable. In contrast, most of the parameter space of Newer General Relativity theory can be ruled out already at the leading order due to the appearance of dangerous extra degrees of freedom\footnote{The situation is similar for New General Relativity \cite{Karananas:2014pxa,Blagojevic:2018dpz} and its generalisations \cite{Koivisto:2018loq}.
yet, we should remark that the absence of pathological degrees of freedom in the linear fluctuations does not guarantee the viability of the theory. In particular, strongly
coupled degrees of freedom seem to be a generic flaw in modified (metric or symmetric) teleparallel gravity theories \cite{Cheng:1988zg,Jimenez:2019ovq,Ferraro:2018tpu,Ferraro:2018axk,Jimenez:2019tkx}.} \cite{BeltranJimenez:2017tkd,Conroy:2017yln}. It could be interesting to study further the novel class of theories that cannot (at least in any straightforward way) be derived from a Lagrangian. At a nonlinear order one should take into account also the potential constitutive relation, which in the general metric case includes 7 additional parameters.

From the above system we can confirm that when $b_1=-b_2=b_3=1$, we have $\nabla_\alpha H^{\alpha\mu}{}_\nu=2\nabla_\alpha {P}^{\alpha\mu}{}_\nu=\tau^\mu{}_\nu$ (setting
$2c_1=-2c_3=-c_2=c_5=-1/2$), where
\begin{subequations}
\ba
\tau^0{}_0 &  = & -2\nabla^2\varphi\,, \\
\tau^0{}_i & = &   -2\dot{\varphi}_{,i} + \frac{1}{2}\nabla^2 V_i\,,\\
\tau^i{}_j  & = &  \lp -\nabla^2\phi + 2\ddot{\varphi} +\nabla^2\varphi + \nabla^2\dot{\beta}\rp\delta^i_j +\lp-\phi+\varphi - \dot{\beta} + \frac{1}{2}\ddot{\sigma}\rp{}^{,i}{}_{,j} - \dot{V}^{(i}{}_{,j)} + \ddot{h}^i{}_j - \nabla^2h^i{}_j\,.
\ea
\end{subequations}
Here, $\varphi$ is the shorthand for $\varphi=\psi-\frac{1}{6}\nabla^2\sigma$.

\subsection{Covariant conservation}

In theories that have a Lagrangian formulation, the concept of conservation is well understood. If the matter couples only to the metric and no other gravitational fields (in particular, a connection with torsion), and we assume a diffeomorphism invariant matter action, then the matter energy-momentum will satisfy the usual metric-covariant conservation law. Even if matter does couple to other gravitational fields, a generalised conservation law will hold, which can be derived in just the same way from diffeomorphism invariance, by looking at a variation $\delta\Phi = \mathcal{L}_{\bxi} \Phi$ of all gravitational fields $\Phi$ given by the Lie derivative with respect to an arbitrary vector field $\bxi$. Then, if the dynamcs of the gravity theory is also described by a diffeomorphism invariant action, it will satisfy an equivalent of the Bianchi identities, and the gravitational field equations relate these generalised Bianchi identities and the conservation of matter energy-momentum \cite{Koivisto:2005yk}.

Since only one special case of the 14-parameter theory studied above admits a Lagrangian formulation, the issue of conservation is a crucial one and needs to be properly addressed. Now we should understand that the consistency condition for the covariant conservation of matter, in particular $\mathcal{D}_\mu T^\mu{}_\nu =0$, determines the equation of motion for the connection. Thus, we still assume that the matter sector of the theory has a Lagrangian formulation - or, at least, that the matter fields obey the usual metric-covariant conservation law and diffeomorphism invariance is valid in this effective sense. It would be possible to consider some more general situation, but that would be non-canonical (as arbitrary prescriptions would be required, e.g. for the connection equation of motion) and not in line with the principles of our axiomatic approach (where the starting point is conservation).

We begin with the field equation for the metric in the 14-parameter theory,
\be \label{pre-efe}
\nabla_\alpha H^{\alpha\mu}{}_\nu = T^\mu{}_\nu + P^\mu{}_{\alpha\beta}Q_\nu{}^{\alpha\beta} - \frac{1}{2}\delta^\mu_\nu P^\gamma{}_{\alpha\beta}Q_\gamma{}^{\alpha\beta}\,.
\ee
From this we derive the connection equation of motion by studying the matter conservation
\be \label{ccoem}
\mathcal{D}_\mu T^\mu{}_\nu = \mathcal{D}_\mu\nabla_\alpha H^{\alpha\mu}{}_\nu - \mathcal{D}_\mu \lp P^\mu{}_{\alpha\beta}Q_\nu{}^{\alpha\beta}\rp + \frac{1}{2}\partial_\nu\lp P^\gamma{}_{\alpha\beta}Q_\gamma{}^{\alpha\beta}\rp\,.
\ee
The metric-covariant derivatives are easily rewritten in terms of our commuting derivatives by noting that for any $(1,1)$-covariant tensor density $X^\mu{}_\nu$ we have (for any torsion-free connection, in fact)
\be \label{mcova}
\mathcal{D}_\mu X^\mu{}_\nu = \nabla_\mu X^\mu{}_\nu + L^\alpha{}_{\mu\nu}X^\mu{}_\alpha - L^\mu{}_{\mu\alpha}X^\alpha{}_\nu - \frac{1}{2}Q_\mu X^\mu{}_\nu
= \nabla_\mu X^\mu{}_\nu + L^\alpha{}_{\mu\nu}X^\mu{}_\nu \,,
\ee
where in the second equality we have taken into account that $L^\mu{}_{\mu\alpha}=-\frac{1}{2}Q_\alpha$, since
\be \label{disformation}
L^\alpha{}_{\mu\nu} = \frac{1}{2}Q^\alpha{}_{\mu\nu} - Q_{(\mu}{}^\alpha{}_{\nu)}\,.
\ee
Thus the first term in (\ref{ccoem}) is very simple,
\be
\mathcal{D}_\mu\nabla_\alpha H^{\alpha\mu}{}_\nu = L^\beta{}_{\mu\nu}\nabla_\alpha H^{\alpha\mu}{}_\beta\,.
\ee
However, in the following it is useful to rewrite this as
\be \label{part1}
\mathcal{D}_\mu\nabla_\alpha H^{\alpha\mu}{}_\nu = -2 L^\beta{}_{\mu\nu} \nabla_\alpha P^{\alpha\mu}{}_\beta + \Delta_\nu =
 -2L_{\beta\mu\nu} \nabla_\alpha P^{\alpha\mu\beta}  - 2L^\alpha{}_{\beta\nu} Q_{\mu\alpha\gamma}P^{\mu\beta\gamma} + \Delta_\nu\,.
\ee
In the first equality we replaced derivative of the kinetic excitation with derivative of the potential excitation, denoting the difference of the corresponding terms as
\be  \label{deltanu}
\Delta_\nu = L^\beta{}_{\mu\nu} \nabla_\alpha\lp H^{\alpha\mu}{}_\beta + 2P^{\alpha\mu}{}_\beta\rp\,,
\ee
and in the second equality we just raised the last index of the tensor inside the derivative. The second term in (\ref{ccoem}) can be expanded into three pieces, using again (\ref{mcova}),
\be \label{part2}
\mathcal{D}_\mu \lp P^\mu{}_{\alpha\beta}Q_\nu{}^{\alpha\beta}\rp = \lp\nabla_\mu P^{\mu\alpha\beta}\rp Q_{\nu\alpha\beta} + P^{\mu\alpha\beta}\lp \nabla_\mu Q_{\nu\alpha\beta}\rp
+ L^\lambda{}_{\mu\nu} P^\mu{}_{\alpha\beta}Q_\lambda{}^{\alpha\beta}\,.
\ee
Now we note that the first term in (\ref{part1}) and (\ref{part2}) enter into the conservation equation (\ref{ccoem}) in the combination
\be
\lp \nabla_\mu P^{\mu\alpha\beta} \rp\lp Q_{\nu\alpha\beta} + 2L_{\alpha\beta\nu}\rp = 0\,,
\ee
because $L_{(\alpha\beta)\nu}=-\frac{1}{2}Q_\nu$, as seen from (\ref{disformation}). We may thus drop those two terms. Let us then consider the remaining 3 terms (forgetting
$\Delta_\nu$ for the moment) we obtain by substracting (\ref{part2}) from (\ref{part1}). By mere index rearranging, we can sum those 3 terms together and obtain
\be
-P^{\mu\alpha\beta}\lp \nabla_\mu Q_{\nu\alpha\beta} + 2L^{\gamma}{}_{\alpha\nu}Q_{\mu\beta\gamma} + L^{\gamma}{}_{\alpha\nu}Q_{\mu\beta\gamma}\rp
= -P^{\mu\alpha\beta}\mathcal{D}_\nu Q_{\mu\alpha\beta} = -\frac{1}{2}\partial_\nu \lp P^{\mu\alpha\beta} Q_{\mu\alpha\beta}\rp\,.
\ee
In the second step we used the property of the commuting covariant derivative that $\nabla_{[\mu}Q_{\nu]\alpha}{}^{\beta}=0$, and then identified the metric-covariant derivative in analogy with the formula (\ref{mcova}). The third step
follows from basic properties of the metric-covariant derivative. The result neatly cancels the remaining piece in (\ref{ccoem}), and thus we have finally arrived at
\be
\mathcal{D}_\mu T^\mu{}_\nu = \Delta_\nu\,.
\ee
This establishes that the equation of motion for the connection in Premetric Newer General Relativity is given by $\Delta_\nu=0$, where $\Delta_\nu$ was defined in (\ref{deltanu}).
Recall that the equation in Newer General Relativity is given by $\nabla_\mu\nabla_\alpha P^{\alpha\mu}{}_\nu=0$. CGR is the singular case that belongs to the union of those two classes of theories, and it is also the unique theory within either class  wherein the equation of motion for the connection trivialises.

Finally, we should note that imposing the $\Delta_\nu=0$ may change the conclusions of the three previous subsections, since properties of the conservative versions of the
14-parameter non-conservative theories can exhibit differences already at the linear order.

\begin{figure}[h]

\begin{tikzpicture}[every text node part/.style={align=center}]
\node[ellipse,draw,label=above:{charge}] (nE) {$E_a$};
\node[ellipse,draw,label=above:{current: matter}](nJ)  [right=of nE,xshift=2cm] {$\bJ_a=\bT_a$};
\draw[->] (nJ)-- node [above,midway] {$\iiint \bJ_a$} (nE);
\node[ellipse,draw,label=above:{gravity}](nT)  [right=of nJ] {$\bt^a$};
\draw[-] (nJ)-- node [above,midway,yshift=0.15cm] {$+$} (nT);

\node[ellipse,draw,label=above:{conjugate}](nX)  [left=of nE,xshift=-2cm] {$x^a = X^{ab}E_b$};
\draw[->] (nE) -- node [above,midway] {$\lb x^a, E_b\rb = i\hbar\delta^a_b$} node [below,midway] {?} (nX);
\node[ellipse,draw,label=below:{kinetic \\ exc.}] [below=of nJ] (nH) { $\bH_a$};
\node[ellipse,draw,label=below:{mass exc.},label=right:{$\Rightarrow$ symmetry \\ breaking}] [below=of nH] (nP) { $\bP_{ab}$};
\draw[<-] (nJ)--  (nH);
\draw[very thin][<-] (nT)--  (nP);
\node[ellipse,draw,label=left:{field strength}](nF)  [below=of nX] {$\bF^{ab}$};
\draw[<-] (nT)--  (nP);
\draw[->] (nX)--  (nF);
\node[rectangle,draw,label=above:{integrability \\ ${\text{\tiny{breaks at E$\sim$M?}}}$}]  (n2)  [right=of nF,xshift=2cm] {$\bF^{ab}=0$};
\draw[dashed][-] (nF)--  (n2);
\node[rectangle,draw,label=above:{field eqn}](n1)  [left=of nH,xshift=-1cm] {$\diff\bH_a=\bJ_a$};
\draw[dashed][-] (nH)--  (n1);
\draw[dashed][-] (nJ)--  (n1);
\node[ellipse,draw,label=below:{potential}](nA)  [below=of nF] {$\bA^{ab}$};
\draw[->] (nF)--  (nA);
\draw[dashed][-] (nA)--  (n2);
\node[ellipse,draw,label=below:{premetric field}](nB)  [below=of n2] {$\diff B^{ab}$};
\draw[->] (nA)-- node [above,midway] {$=$} (nB);
\draw[->] (n2)--  (nB);
\draw[very thick][->] (nB) -- node [above,midway] (c1) {$\chi$}  (nH);
\draw[very thick][->] (nB) --  node [above,midway] (c2) {$\xi$} (nP);
\draw[<->]   (c1) [xshift=10cm,yshift=-15cm] --  node [right,midway] {$\exists \bL  \Rightarrow \underset{\text{\tiny{CGR}}}{(\chi,\xi)}$}  (c2);  
\node[ellipse,label=below:{}](nS)  [right=of nP] {$$};
\node[ellipse,label=left:{scale}](nM)  [below=of nT,yshift=-0.5cm] {$M$};
\draw[->] (nS)--  (nM);
\draw[->] (nM)--  (nT);

\end{tikzpicture}

\caption{A schematic figure illustrating the logical structure of the premetric construction of purified gravity. The constitutive relations are $\chi$ and $\xi$ (with indices omitted).
The kinetic excitation is related to the existence of a conserved current, and the mass excitation is related to the presence of gravitational contribution to the current. At some level, the energy and momenta are conjugate to space and time. The way we set up the coordinates for the latter (or, the frame), is merely a convention. The choice (which may become, even in principle, impossible at the Planck scale) will affect our description of physics, but this effect has to be purely inertial and not a physical force. The gauge potential is thus given by a gauge transformation. Due to the nonzero mass excitation the gauge transformation becomes the dynamical Stueckelberg field. Thus, the constitutive relation $\xi$ renders the $B^{ab}$ into a dynamical ``premetric field'', and in symmetrised conjuction with the constitutive relation $\chi$ filters from the field the properties of a metric in the unique fashion that is dictated by the requirement of an underlying action principle. \label{fig1}}

\end{figure}

\section{Conclusions and perspectives}
\label{conclusions}

The conclusion of this paper is given by the Figure \ref{fig1} and the formula (\ref{two}), where the former illustrates the axiomatic deduction of the CGR purified gravity theory, and the
latter specifies its suggested extrapolation. In what follows we will discuss these conclusions at more length.

\subsection{Summary}

\subsubsection{Fundamental equations}

In the premetric program one foundational dichotomy is the separation of extensive (how many) and intensive (how strong) quantities.
Our starting point for the extensive objects, in the matter sector, was the conservation of energy and momenta, in particular, a 4-component conserved charge. For the intensive objects, in the gravitational sector, the main assumption was the integrability postulate, in particular, the vanishing of the field strength of a 16-component one-form potential. From this we arrived at a class of theories that have a very close analogy to the theory of massive electromagnetism (as was exhibited in Table \ref{table1}). The relation that was deduced between the extensive quantities,
\begin{itemize}
\item $\text{Fundamental Equation 1:} \quad \nabla_\alpha H^{\alpha\mu}{}_\nu = T^\mu{}_\nu + t^\mu{}_\nu$\,, \label{field_eq}
\end{itemize}
where now the presence of $t^\mu{}_\nu$ is the consequence of the symmetry breaking realised by the mass of the one-form potential, is the fundamental equation that in the end determines the dynamics of the fields.
The other fundamental equation is the integrability postulate that now determines the nature of the intensive quantities,
\begin{itemize}
\item $\text{Fundamental Equation 2:} \quad F^{\alpha\beta}{}_{\mu\nu} \approx 0\,.$ \label{integrability}
\end{itemize}
This means that the one-form does not propagate. However, due to the symmetry breaking the theory is not quite trivial. Analogously to pure-gauge Proca theory where the gauge field content reduces to one massless scalar,
in our generalisation of the pure-gauge massive field theory there remains a propagating massless tensor field, the premetric field. This structure of the theory is schematically illustrated in Fig. \ref{fig1}.

\subsubsection{Linking equations}

\begin{figure}
\begin{center}
\begin{tikzpicture}[every text node part/.style={align=center}]
\node[ellipse,draw,label=above:{general},label=left:{$\xi$}] (nG) {$4096$};
\node[ellipse,draw,label=above:{symmetrised}](nS)  [right=of nG] {$1600$};
\draw[->] (nG)--  (nS);
\node[ellipse,draw,label=above:{metric}](nM)  [right=of nS] {$7$};
\draw[->] (nS)--  (nM);
\node[ellipse,draw,label=above:{parity-even}](nE)  [right=of nM] {$6$};
\draw[->] (nM)--  (nE);
\node[ellipse,draw,label=left:{reversible $\xi$}](nG2)  [below=of nG] {$2176$};
\draw[dashed][->] (nG)--  (nG2);
\node[ellipse,draw](nS2)  [right=of nG2] {$820$};
\draw[dashed][->] (nS)--  (nS2);
\draw[->] (nG2)--  (nS2);
\node[ellipse,draw](nM2)  [right=of nS2] {$6$};
\draw[dashed][->] (nM)--  (nM2);
\draw[->] (nS2)--  (nM2);
\node[ellipse,draw](nE2)  [right=of nM2] {$5$};
\draw[dashed][->] (nE)--  (nE2);
\draw[->] (nM2)--  (nE2);
\node[ellipse,draw,label=above:{compatible}](nC)  [right=of nE2] {$1$};
\draw[->] (nE)--  (nC);
\draw[->] (nE2)--  (nC);
\node[ellipse,draw,label=left:{$\chi$}](nGc)  [below=of nG2] {$1536$};
\node[ellipse,draw](nSc)  [below=of nS2] {$960$};
\node[ellipse,draw](nMc)  [below=of nM2] {$7$};
\node[ellipse,draw](nEc)  [below=of nE2] {$3$};
\draw[->] (nGc)--  (nSc);
\draw[->] (nSc)--  (nMc);
\draw[->] (nMc)--  (nEc);
\draw[->] (nEc)--  (nC);
\end{tikzpicture}
\end{center}
\caption{The number of independent components in the constitutive relations in cases satisfying certain conditions. The solid lines indicate restrictive assumptions on the constitutive relations,
and the dashed lines in particular indicate the assumption of reversibility of the potential constitutive relation. E.g., in the generic symmetrised, there are in total 1560 independent components that can determine
the theory. In the case that the constitutive relations are given in terms of a metric, the number reduces to 13, and the further requirement of an action principle leaves no freedom expect for an overall constant. \label{fig2}}
\end{figure}
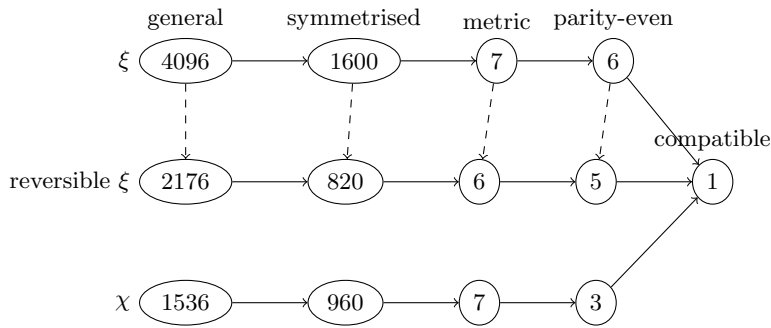

At this stage the theory was completely metric-free. It was also non-predictive, since the quantities appearing in the would-be dynamical Fundamental Equation 1 are undetermined. The theory is completed by establishing the
relations that link the extensive and the intensive quantities, called the constitutive relations. The intensive quantity in our case is the premetric field $B^{\mu\nu}$, and due to its Stueckelbergian origin it appears, at the pre-metric level,
only via its derivatives. Assuming a linear constitutive relation, the kinetic excitation appearing in the Fundamental Equation 1 is given as
\begin{itemize}
\item $\text{Linking Equation 1:} \quad H^{\alpha\mu}{}_\nu = \chi^{\alpha\mu}{}_{\rho\sigma}{}^{\beta}\nabla_\beta  B^{\rho\sigma}\,.$ \label{link1}
\end{itemize}
The constitutive tensor is antisymmetric in its first indices, otherwise in principle arbitrary. In this study we restricted to relations which also are symmetric in the last two covariant indices. This way the desired metric properties are inherited by the premetric field. We considered two irreducible decompositions of the general relation, separating it into four and five irreducible
components, respectively. They determine the piece $t^\mu{}_\nu$  in the Fundamental Equation 1, whose form is given by (\ref{emt}), together with the potential constitutive relation,
\begin{itemize}
\item $\text{Linking Equation 2:} \quad P^{\alpha}{}_{\mu\nu} = \xi^{\alpha}{}_{\mu\nu}{}^{\beta}{}_{\rho\sigma}\nabla_\beta B^{\rho\sigma}\,.$ \label{link2}
\end{itemize}
Again, we restricted ourselves to symmetrised relations, such that the two pairs of covariant indices are symmetrised. We then recognised six irreducible components of the relation: symmetric and antisymmetric principal
components, which are both reversible; symmetric and antisymmetric skewon components, which are both irreversible; and an axion component which is further reduced to the reversible part and the irreversible part; recall Table \ref{nomenclature}.
The total number of independent components, assuming different conditions on the constitutive relations, is reviewed in Figure \ref{fig2}.

\subsubsection{Metric purified gravity}
\label{mpg}

We investigated in more detail the cases where the two Linking Equations involve a metric. In particular, since the field $B^{\mu\nu}$ had emerged from the premetric structure as a consequence of
symmetry breaking, it was the canonical candidate for the role of the metric. It had entered into the theory as a Stueckelberg field restoring the symmetry broken by the mass term in the potential, implied by the
nontrivial Linking Equation 2. The generic metric constitutive relations have 14 independent components, contributing to all but the antisymmetric skewon and the irreversible axion irreducible parts of the relations.
As a first viability check of the new class of theories, which could be dubbed the Premetric Newer General Relativity, we explored the particle content and the wave propagation in the weak-field limit. Preliminarily,
we could exclude only two components as they contribute to the antisymmetric parts of the Fundamental Equation 1 and would result in new, and probably dangerous, degrees of freedom. Without additional
constraints, the general theories would also violate the equivalence principle, in the sense that matter would not follow the metric-geodesic trajectories. In Lagrangian theories, the conservation of matter follows from
the diffeomorphism invariance of the action. However, in our premetric construction we do not presuppose a Lagrangian formulation. Therefore an additional constraint may be in order,
\begin{itemize}
\item $\text{Geodesic Postulate:} \quad \mathcal{D}_\mu T^\mu{}_\nu=0\,.$
\end{itemize}
Indeed, even in the context of General Relativity, the geodesic motion of matter is sometimes introduced as an independent postulate. However, in Lagrangian theories the laws governing the motion of particles are
inscribed in the field equations. In the context of Lagrangian symmetric teleparallel theory, the metric-covariant conservation of matter energy-momentum follows from the extremisation of the action with respect to the
variations of connection. CGR is the unique quadratic theory whose action is extremised by an arbitrary connection, meaning that {\it the Geodesic Postulate is redundant with the Fundamental Equation 1}. Regardless
of this, in principle, the CGR is also uniquely specified, amongst the 14 possible metric constitutive relations, by requiring that {\it the Linking Equation 1 is compatible with an action principle determined by the Linking Equation 2}. In the more general Premetric Newer General Relativity, the geodesic postulate has to be separately imposed, and it can be regarded as the equation of motion for the symmetric teleparallel connection that cannot now be deduced by the usual variational methods due to the absence of a Lagrangian. In the case of the 14-parameter theory, the equation
is $\Delta_\mu=0$, where $\Delta_\mu$ was defined in (\ref{deltanu}).

\subsection{Comparison with metric teleparallelism}

There are two main points of departure in the premetric construction of purified gravity (PG) in comparison to the premetric construction of metric teleparallel gravity (TG).
1) Firstly, in PG it is imposed the vanishing of the force, while an excitation is allowed. Since in TG there is a force like in electromagnetism, the analogy with Maxwell theory remains intact at this point. On the other hand, the
fact that gravity indeed is different from electromagnetism (and the other interactions) in that it is equivalent to inertia, is incorporated into the premetric structure of PG.
Secondly, 2) PG is analogous to the massive rather
than the vanilla Maxwell electromagnetism. On the other hand, in TG one then needs to break the complete analogy by introducing gravitational charges, which have no  correspondence in the Maxwell electromagnetism.
Related to this point, one could perhaps add that 2') in PG the metric is a Stueckelberg field of the (pure-gauge) connection, while in TG the metric is obtained from the translational gauge connection by identifying it with the coframe field. However, the latter identification entails the tacit introduction of a symmetry-breaking field.

Thus, through the point 2), the necessary breaking of symmetry that has been left unaddressed in the premetric discussions of TG, is raised to a main role in the premetric construction of PG.   
Further, the proposed extrapolation of PG towards the ultra-violet regime, recall (\ref{two}), would restore the one-to-one correspondence with massive electromagnetism that is
only apparently lost by the point 1) at macroscopic scales

\subsection{Implications for CGR}

The premetric approach provides some new insights to CGR, which emerges as the unique metrical theory that is consistent with an action formulation and the axioms of the premetric framework.
In this framework, the field equation is given by the Fundamental Equation 1 above, with the constitutive relations from (\ref{cgrconst}). In particular, it features the tensor density $H^{\alpha\nu}{}_\nu$, which can be shown to reduce, in the coincident gauge, to what is known as the Einstein energy-momentum complex pseudotensor density. It has been found that by using that pseudotensor density one always obtains reasonable results for the energy-momentum of any gravitational-matter system, whereas various other pseudotensorial prescriptions sometimes fail to yield the expected answer\footnote{For comparative studies see e.g. \cite{Aguirregabiria:1995qz,Xulu:2002ix}.}. In CGR the Einstein energy-momentum complex is promoted to a true tensor
density, and the recently proposed characterisation of the canonical frame has removed the ambiguity from the results. However, it was left unjustified why the tensor density $\tau^\mu{}_\nu=\nabla_\alpha H^{\alpha\mu}{}_\nu$ should be used in the determination of the gravitational energy-momentum instead of the $\hat{\tau}^\mu{}_\nu=-2\nabla_\alpha P^{\alpha\mu}{}_\nu$, which is more naturally written into the action-derived field equations. The ambiguity is related to the freedom to add a so called superpotential to a given energy-momentum complex. However, the canonical premetric framework leaves no such freedom. By construction it is obvious that we are compelled to use $H^{\alpha\nu}{}_\nu$ to determine the translational charges, for it is precisely the excitation conjugated to the translational currents.

\subsection{Implications for extended gravity theories}

The class of extended gravity theories we studied in more detail in this work was defined by the 14-parameters of the general metric linear (and local) constitutive relation, the Premetric Newer General Relativity.
As reviewed above, the class of models survives the first consistency and viability tests since they (given only two constraints on the parameters) reduce to General Relativity
at the linear order. This can be contrasted with the generic quadratic metric teleparallel and quadratic symmetric teleparallel theories (sometimes referred to the New and the Newer General Relativity, respectively), whose parameters are very stringently constrained in the same limit (and as well known, for quadratic pure-metric theories only the topological Gauss-Bonnet terms survive). It can be therefore interesting to pursue further the newly found theories, to investigate the viability of their nonlinear solutions and for example in view of their possible cosmological applications.

In general, the premetric approach raises the perhaps exotic possibility of theories without a Lagrangian formulation, of which the quadratic 14-parameter class is just an example.

Considering Lagrangian extended theories characterised by more general constitutive relations, there are two main observations we can make. Firstly, many of the previously studied extended symmetric teleparallel theories, in particular such
with nonlinear constitutive relations, might be difficult to incorporate within the premetric formalism. As a simple example, in the $f(Q)$ models one would require an excitation tensor $H^{\mu\nu}{}_\nu$ such that $\nabla_\alpha H^{\alpha\mu}{}_\nu = -2\nabla_\alpha f'(Q)\prescript{}{CGR}P^{\alpha\mu}{}_\nu$, which at first look would not appear easily possible unless $f'(Q)$ is a constant. Looking at things from the other side, the premetric framework suggests a great variety of previously unexplored ways of extending gravity. As an example, as we imposed three symmetrisations upon the constitutive relations from the beginning, mainly motivated by the aim of obtaining a metric in the end, it is reasonable to ask what what would happen when some or all of these symmetrisations were abandoned. This would expand the available theory space, and provide a remarkably simple way of realising asymmetric gravity which from the outset avoids the main problems that there are in extending the purely-metric theory by allowing antisymmetric components in the metric tensor. Namely, in the latter case one quite generally introduces ghosts since the available invariants are of a higher order, and one also encounters technical obstacles e.g. in the generalisation of the Levi-Civita connection\footnote{See e.g. \cite{Mann:1981st,Damour:1992bt,Moffat:1994hv,Janssen:2006jx,Janssen:2007yu} for studies on gravitational and theories with nonsymmetric metric.}.

Another Pandora's box is opened by allowing more extra fields to determine the constitutive relations. One non-minimal but rather natural extension would be to consider the case that the constitutive relation is not determined by the premetric field, but by an independent metric. Of course, this kind of bimetric theory and the many other possible novel extensions could turn out to be plagued by ghosts or other pathologies\footnote{A bimetric constitutive relation to an extent resembles such bimetric variational principles wherein the connection is considered as the Levi-Civita connection of an independent metric \cite{Koivisto:2011vq,Amendola:2010bk}. The latter setups may however introduce problematical degrees of freedom \cite{BeltranJimenez:2012sz}. For a review of the ghost-free bimetric gravity theory \cite{Hassan:2011zd}, see \cite{Schmidt-May:2015vnx}.}. However, one may contemplate whether it is possible to establish a robust correspondence between the field-theoretical consistency of a purified gravity theory and the existence of its
action-compatible premetric formulation. Such quite unique cases of consistent theories as CGR and (the symmetric teleparallel version of) the ghost-free Hassan-Rosen bimetric theory turn out to be also quite unique
in that they admit a premetric formulation.

One of our conclusions at least is that without the latter, a theory does not have a well-defined canonical energy-momentum complex.

\subsection{Implications for quantum gravity}

First we should recall that purified gravity has already provided insights that could be highly relevant in the unification of gravity and quantum physics. In the canonical approach to quantum gravity, the notorious problem of time might be taken into reconsideration from the perspective wherein we, besides the conventional ADM energy of the standard Hamiltonian formalism, have also available the unique consistent definition of localisable energy excitation in a gravitational system. The other main approach to quantisation, the path integral formalism, can obviously also have a more promising starting point, since the CGR action in the canonical frame
 is well-defined without invoking boundary terms or counter terms that are necessary e.g. in the standard approach to Euclidean quantum gravity in Riemannian geometry.

In the study carried out in this paper, the question about whether the analogy of purified gravity with Proca electromagnetism is complete naturally arose, the vanishing of the field strenght tensor then being a valid approximation only at super-Planckian length scales.
Though massive Abelian gauge theories are renormalisable even without the Higgs mechanism, from the perspective of purified gravity it is natural to consider a spontaneous emergence of the Planck scale
since one wants to recover scale invariance at the most fundamental level of physics. In this work we only very tentatively discussed an actual realisation of such a mechanism (in particular, an analogous mechanism with Kibble's
spontaneous mass generation), but the new way of looking at gravity, seeing the covariant version of the Einstein Lagrangian as the mass term for the gauge connection instead of a kinetic term, very concretely points to a quite novel approach to realise gravity as a renormalisable gauge field theory with no formal difference from the others we already know.

That the metric is a part of the connection that is curved only at very microscopic distances, raises speculations about the physical role of that curvature. One could be brought back to pre-Einsteinian considerations of geometric
theories of gravity, in particular to Riemann and Clifford, who both entertained the idea that matter is nothing but a tiny disturbance in the spatial curvature, so that matter in motion can be understood as a simple variation in space
of these disturbances. Now the fact that the metric, a macroscopic emergent field which also can be associated with a curvature at a less fundamental level, is interwined with matter via the Einstein equation, could be understood
as a consequence of both the metric and the matter being aspects of the same connection. In the premetric construction of electromagnetism, one ultimately builds upon quantities that can be counted: electric charges and magnetic fiux
lines. It is clear that energy and momentum are quantised, and thus our starting point of the conservation of energy-momentum is in line with the principle of countability of extensive quantities. Perhaps the countability of intensive
quantities, in this case the ones related to the flux lines of the ``hypergravitational'' field, should be understood as a reflection of the quantised nature of matter particles.


 \acknowledgements{The work was supported by the Estonian Research Council through the Personal Research Funding project PRG356 ``Gauge Gravity'' and by the European Regional Development Fund through the Center of Excellence TK133 ``The Dark Side of the Universe''.}

\bibliography{PreQ}

\end{document}